\documentclass[a4paper,12pt]{article}
\usepackage{graphicx} 
\usepackage{geometry}
\usepackage{amsmath,amssymb,amsfonts}
\usepackage{xcolor}
\usepackage{subcaption}
\usepackage{url}
\usepackage{slashed} 
\usepackage{enumerate}
\usepackage{epsfig} 
\usepackage{float}
\usepackage{jheppub}
\usepackage{graphicx}

\usepackage[utf8]{inputenc}  

\title{Determination of $\mathcal{CP}$-violating $HZZ$ interaction with polarised beams at the ILC}
\author[1]{Cheng Li}
\author[2,3]{Gudrid Moortgat-Pick}
  \affiliation[1]{School of Science, Sun Yat-Sen University, Gongchang Road 66, 518107 Shenzhen, China}
  \affiliation[2]{II. Institut f{\"u}r Theoretische Physik\\  Universit{\"a}t Hamburg, Luruper Chaussee 149, 22761 Hamburg, Germany}

 \affiliation[3]{Deutsches Elektronen-Synchrotron DESY,
  Notkestr. 85, 22607 Hamburg, Germany}

 \emailAdd{cheng.li@desy.de}
  \emailAdd{lich389@mail.sysu.edu.cn}
  \emailAdd{gudrid.moortgat-pick@desy.de}
 
\abstract{

We study possible $\cp$-violation effects of the 125~GeV Higgs to $Z$ boson coupling at the 250~GeV ILC with transverse and longitudinal beam polarisation via the process $e^+ e^- \rightarrow HZ \rightarrow H \mu^-\mu^+$. We explore the azimuthal angular distribution of the muon pair from the $Z$ boson decay, and constructe $\cp$-odd observables sensitive to $\cp$-violation effects, where we derived this observable both by analytical calculations and by \texttt{Whizard} simulations. Particularly, we can construct two $\cp$-odd observables with the help of transversely-polarised initial beams and improve the statistical significance of $\cp$-violation effects by combining two measurements. We defined the asymmetries between the signal regions with different signs of the $\cp$-odd observables, and determine the $\cp$-violation effect by comparing with the SM 95\% C.L. upper bound. In this paper, we setup a scenario which assumes that the total cross-section is always fixed while $\cp$-violation is varying, and such a scenario helps us to determine the intrinsic $\cp$-mixing angle limit around $|\xi_\cp|\sim 0.03$ with (90\%, 40\%) polarised electron-positron beams and 5~ab$^{-1}$ integrated luminosity. In addition, we determine the $\cp$-odd coupling limit $|\widetilde{c}_{HZZ}|\sim 0.01$ as well, where we suppose that the SM tree-level cross-section is fixed and the $\cp$-violation is the varying additional contribution. Comparing with the analysis with unpolarised beams, the sensitivity to the $\cp$-violation effect can be improved by transverse or longitudinal polarisation.

}
\usepackage{abbrev}

\begin{document}

\maketitle

\section{Introduction}
\label{sec:intro}

Since 2012, a Higgs boson with a mass of 125~GeV has been discovered by both ATLAS and CMS collaborations \cite{ATLAS:2012yve,CMS:2012qbp} within the experimental and theoretical uncertainties, that are consistent with the expectations in the Standard Model (SM) of elementary particle physics. So far, the LHC experiment has not discovered significant evidence for physics beyond the Standard Model (BSM). However, the measured Cosmic Microwave Background anisotropies \cite{Planck:2018vyg} demonstrate, for instance, that the Universe has a much larger baryon-antibaryon asymmetries than the SM predictions can embed. In principle, the occurrence of Baryogenesis in the early universe acquires the Sakharov conditions \cite{Sakharov:1967dj}, which cannot be fulfilled in the SM. Therefore, the Two-Higgs-Doublet Model (2HDM) \cite{Lee:1973iz} with a complex vacuum expectation value (vev), called complex 2HDM (C2HDM), is motivated to introduce an additional source of $\cp$-violation and can accommodate the required strong first-order phase-transition. 
In the C2HDM \cite{Gunion:2005ja}, the 125~GeV Higgs boson is an admixture of scalar and pseudoscalar components, and the process with Higgs to fermions interaction are $\cp$-violating:
\begin{equation}
    \mathcal{L} \supset \Bar{f}(c_{Hf\Bar{f}} + i\gamma_5 \tilde{c}_{H f\bar{f}})f H.
\end{equation}
Hence, the $\cp$ structure of the $Ht\bar{t}$ interaction and the impact on Baryogenesis has been exploited by applying LHC searches via $t\bar{t}H$ production \cite{ATLAS:2019nvo,ATLAS:2020pvn, ATLAS:2020ior,CMS:2019lcn,CMS:2020cga}, and the results are summarized in \cite{Bahl:2020wee,Gritsan:2022php, Dawson:2022zbb}, while the effect of electron EDM and baryogenesis in incorporated and discussed together with the Higgs $\cp$ structure measurement in \cite{Bahl:2022yrs}.

At the tree level within the C2HDM, the Higgs to gauge boson $HVV$ interactions are still $\cp$-conserving. However, the $\cp$-violating Higgs to fermion couplings can change the $\cp$ structure of the $HVV$ interactions at the one-loop level, where the imaginary part of the $Hff$ couplings leads to the $\cp$-odd part of $HVV$ interactions. This is called anomalous Higgs to gauge-bosons coupling and shown in the following
\begin{equation}
    \mathcal{L}\supset \frac{\widetilde{c}_{HVV}}{\Lambda} H V_{\mu\nu} \widetilde{V}^{\mu\nu} ,
\end{equation}
where $\Lambda$ is the new physics scale and 
\begin{equation}
       V_{\mu\nu}=\partial_\mu V_\nu -\partial_\nu V_\mu,\qquad \widetilde{V}_{\mu\nu} =\frac{1}{2}\epsilon_{\mu\nu\rho\sigma}V^{\rho\sigma}.
\end{equation}
Therefore, collider phenomenology of the $\cp$ structure of the Higgs to gauge-bosons interaction can be investigated, and the LHC has performed the corresponding searches via the VBF and VH productions and $H\rightarrow 4\ell$ decay at CMS \cite{CMS:2017len,CMS:2019jdw,CMS:2019ekd,CMS:2021nnc,CMS:2022uox,CMS:2024bua} and at ATLAS \cite{ATLAS:2020evk,ATLAS:2022tan,ATLAS:2023mqy}. So far, the latest LHC experiments provide the observed limits of CP-odd $HVV$ coupling, which are $\left({\tilde{c}_{ZZ}} \right)_\text{CMS} \sim [-0.66, 0.51]$ \cite{CMS:2024bua} and $(\tilde{c}_{ZZ})_\text{ATLAS}\sim [-1.2, 1.75]$ \cite{ATLAS:2023mqy} at 95\% C.L.. 

Furthermore, the study of $\cp$ properties of the Higgs boson can also be performed at future colliders. The HL-LHC study \cite{Cepeda:2019klc} provides the prospect of future measurement with 3~ab$^{-1}$. The electron-positron colliders are very promising, where the CPEC [ref] can provide the unpolarised electron-positron beams at 240~GeV with 5.6~ab$^{-1}$ and 20~ab$^{-1}$. Based on the CEPC setup, the $\cp$-violation effect on $H\tau^+\tau^-$ interaction has already been studied via $H\rightarrow\tau\tau$ decay \cite{Ge:2020mcl}, while the studies of \cite{Jovin:2021qnz,Jeans:2018anq} investigated the $\cp$-violating $H\rightarrow\tau^+\tau^-$ decay at the ILC as well. On the other hand, the recoil $Z$ boson from the Higgs strahlung at $e^+ e^-$ colliders can also carry information of the $HZZ$ interaction, and one can study the $\cp$ structure by the recoil $Z$ decays. Therefore, the study of \cite{Sha:2022bkt} performs the determination of $HZZ$ coupling via the $e^+ e^-\rightarrow H Z\rightarrow H \ell^- \ell^+$ at the CEPC, where the initial beams are currently foreseen to be unpolarised. Besides, CLIC also provides the studies of the $\cp$-structure of $HVV$ coupling \cite{Karadeniz:2019upm,Vukasinovic:2023jxd}, where the vector-boson-fusion would be the dominant process at above 1~TeV.
However, the ILC could generate simultaneously polarised electron and positron beams, so that also transversely or longitudinally polarised beams (provided by applying spin rotators) can be exploited for the analysis \cite{Moortgat-Pick:2005jsx}. By using this initially polarised beams, the sensitivity to the $\cp$ violation effect can be potentially improved compared to the case without beam polarisation. Thus, one can use transversely-polarised beams to test $\cp$-violation effects in the $e^+ e^- \rightarrow H Z$ process, which is already proposed by the studies \cite{Rao:2007ce,Biswal:2009ar,Rindani:2010pi}, and can provide the future aspects of the determination of the $\cp$-violation coupling.

In this work, we focus on the Higgs strahlung process at the ILC with a center of mass energy of 250 GeV, apply transversely or longitudinally polarised electron-positron beams, calculate the scattering amplitude analytically and obtain the cross-section by numerical integration. 
Based on the analysis of the azimuthal angular distribution of the muon pair produced by the $Z$ decay, we construct T-odd observables to probe the CP-violation effect. Particularly, we can define two $\cp$-odd observables when the transverse polarisation is imposed, where one of the additional observable is defined by the spin orientation of electron-positron beams. Therefore, we perform the Monte-Carlo simulation by \texttt{whizard-3.0.3} \cite{Kilian:2007gr,Moretti:2001zz}, and obtain the number of events in the corresponding signal regions with different sign of the $\cp$-odd observables. These number of events can be used to construct the asymmetries, as well as carrying out the likelihood fit, to determine the size of the $\cp$-violation effect. We setup two scenarios for the determination, where the first scenario consists of the fixing total cross-section for varying intrinsic $\cp$-mixing angle, and we can determine the $\cp$-mixing angle $|\xi_\cp|\sim 0.035$ with 5~ab$^{-1}$ integrated luminosity, where the initial beams are (90\%, 40\%) transversely polarised. 
However, the longitudinal polarisation can enhance the total cross-section and suppresses the statistical uncertainty, leading to $|\xi_\cp|\sim 0.03$ with (-90\%, 40\%) polarisation degrees and 5~ab$^{-1}$. For the second scenario, we can fix the SM tree-level contribution and vary the additive $\cp$-odd contribution. In this second scenario, we determine the $\cp$-odd coupling $|\widetilde{c}_{HZZ}|\sim 0.01$ with 5~ab$^{-1}$, where both (90\%, 40\%) transverse and longitudinal polarisation lead complementary to similar results of determination.


\section{The $\cp$-violation in the Higgs boson}

\label{sec:higgscharac}

In general, we can apply the Higgs characterization model for the 125~GeV Higgs boson~\cite{Artoisenet:2013puc}, which is effective approach for all possible 125~GeV Higgs boson interactions, without introducing irrelevant higher dimension operators. The effective Lagrangian of the Higgs characterization model is given by:
\begin{multline}
    \mathcal{L}_\text{eff}=\Big[    \cos\xi_{\cp}~ \kappa_\text{SM}(\frac{1}{2}g_{HZZ} Z_\mu Z^\mu +g_{HWW} W^+_\mu {W^-}^\mu)\\
    -\frac{1}{4}(\cos\xi_{\cp}~\kappa_{H\gamma\gamma}A_{\mu\nu}A^{\mu\nu}+\sin\xi_{\cp}~\widetilde{\kappa}_{H\gamma\gamma}A_{\mu\nu}\widetilde{A}^{\mu\nu}) \\
    -\frac{1}{2}(\cos\xi_{\cp}~\kappa_{H Z\gamma}Z_{\mu\nu}A^{\mu\nu}+\sin\xi_{\cp}~\widetilde{\kappa}_{HZ\gamma}Z_{\mu\nu}\widetilde{A}^{\mu\nu})\\  
    -\frac{1}{4}(\cos\xi_{\cp}~\kappa_{Hgg}G_{\mu\nu}G^{\mu\nu}+\sin\xi_{\cp}~\widetilde{\kappa}_{Hgg}G_{\mu\nu}\widetilde{G}^{\mu\nu})\\
    -\frac{1}{4\Lambda}(\cos\xi_{\cp}~\kappa_{HZZ}Z_{\mu\nu}Z^{\mu\nu}+\sin\xi_{\cp}~\widetilde{\kappa}_{HZZ}Z_{\mu\nu}\widetilde{Z}^{\mu\nu}) \\ 
    -\frac{1}{2\Lambda}(\cos\xi_{\cp}~\kappa_{HWW}W^+_{\mu\nu}{W^-}^{\mu\nu}+\sin\xi_{\cp}~\widetilde{\kappa}_{HWW}W^+_{\mu\nu}{\widetilde{W}}^{-\mu\nu})\\
    -\frac{\cos\xi_{\cp}}{\Lambda}(\kappa_{H\partial\gamma}Z_\nu\partial_\mu A^{\mu\nu}+\kappa_{H\partial Z}Z_\nu\partial_\mu Z^{\mu\nu}+(\kappa_{H\partial W} W^+_\nu\partial_\mu {W^-}^{\mu\nu}+\text{h.c.}))\Big]H\\
    -\sum_{f}\Bar{f}(\cos\xi_{\cp}~ c_{Hff} + i\sin\xi_{\cp}~ \widetilde{c}_{Hff}\gamma_5)f H,
    \label{eq:Heffl}
\end{multline}
where the $A,Z,W^\pm,G$ are the photon, $Z$ boson, $W$ boson and gluon fields respectively, and $\Lambda$ is the new physics scale of effective field theory. This model contains all the possible Higgs interactions to the other SM particles. In the effective Lagrangian, the parameter $\xi_{\cp}$ is the $\mathcal{CP}$-mixing angle of the Higgs boson, so that $\xi_{\cp}\neq0$, $\xi_{\cp}\neq \pm\frac{\pi}{2}$ and non-zero $\widetilde{c}_{Hff},\widetilde{\kappa}_{HVV}$ imply $\mathcal{CP}$ violation. 
Particularly, we focus on the $HZZ$ couplings, which contribute via the following terms
\begin{equation}
    \mathcal{L}\supset c_\text{SM} \frac{m_Z^2  }{v}  Z_\mu Z^\mu H -\frac{c_{HZZ}}{v} Z_{\mu\nu}Z^{\mu\nu}H -\frac{\widetilde{c}_{HZZ}}{v}Z_{\mu\nu}\widetilde{Z}^{\mu\nu}H,
    \label{eq:effL}
\end{equation}
where:
\begin{align}
   &c_\mathrm{SM}=\kappa_\text{SM} \cos\xi_{\cp},\\
    &\widetilde{c}_{HZZ} = \frac{1}{4}\widetilde{\kappa}_{HZZ}\sin\xi_{\cp} ,\\
    &c_{HZZ} = \frac{1}{4}{\kappa}_{HZZ}{\cos\xi_{\cp} }.\label{eq:dim6hzz}
\end{align}
Since we are interested in the physics at Electroweak scale, we choose $\Lambda = v\approx 246$~GeV. The coefficients $c_\mathrm{SM}$, $c_{HZZ}$ and $\widetilde{c}_{HZZ}$ parameterize all the possible contributions to the corresponding operators. In an UV complete model (e.g. C2HDM), the coefficients of the one-loop contribution $\kappa_{HZZ}$ and $\widetilde{\kappa}_{HZZ}$ can be solved by summing up all the loop integrals. However, these couplings can be suppressed by the factor $\frac{1}{16\pi^2}$, while the experimental constraints on these couplings of 125~GeV Higgs are relatively loose. In this case, this $\cp$-odd term of $HZZ$ interaction may be contributed by other sources.

For the scattering process with one $HZZ$ vertex, the scattering amplitude can be evaluated by:
\begin{equation}
\begin{split}
    \mathcal{M}\propto&\frac{\mathcal{M}^{\mu\nu }}{v}\left[{ c_\mathrm{SM}} m_Z^2  \, g_{\mu\nu} + {c_{HZZ}}(q_{1\nu}q_{2\mu}-g_{\mu\nu}q_1\cdot q_2) +{\widetilde{c}_{HZZ}}\,\epsilon_{\mu\nu\alpha\beta}~{q_1^\alpha q_2^\beta}\right]\\
&=\frac{\mathcal{M}^{\mu\nu }}{v}\Big[\cos\xi_\cp \left( \kappa_\mathrm{SM} m_Z^2 \, g_{\mu\nu} + \frac{\kappa_{HZZ}}{4}(q_{1\nu}q_{2\mu}-g_{\mu\nu}q_1\cdot q_2)\right) \\&\hspace{6cm}+\sin\xi_\cp\frac{\widetilde{\kappa}_{HZZ}}{4} \epsilon_{\mu\nu\alpha\beta}~{q_1^\alpha q_2^\beta}\Big],
    \label{eq:cpmixhvv}
\end{split}
\end{equation}
where the momenta $q_1$ and $q_2$ are the momenta of the $Z$ bosons (see Fig.~\ref{fig:diageezh}). In this amplitude, the SM tree-level term with $c_\mathrm{SM}$ and the next-to-leading-order term with ${c_{HZZ}}$ are both $\cp$-even, while the term with ${\widetilde{c}_{HZZ}}$ is the leading-order $\cp$-odd term.

\section{The production and decay process at the ILC}
\subsection{The initial polarised electron-positron beams}
\label{sec:pol}
Concerning the polarisation of the initial electron and positron beams, one can define a projection operator, that is called polarisation matrix:
\begin{equation}
    \frac{1}{2}(1 - \boldsymbol{P}\cdot \boldsymbol{\sigma}) =\frac{1}{2}(\delta_{\lambda\lambda'}-P^a\sigma^a_{\lambda\lambda'}) = \frac{1}{2}\begin{pmatrix}
        1- P^3& P^1- iP^2\\
        P^1 + iP^2& 1+P^3
    \end{pmatrix},
    \label{eq:polmat}
\end{equation}
where the $\boldsymbol{P}$ is the polarisation vector of the electron beam. More explicitly, the polarisation vector can be parameterised by the polarisation fraction $f$ and the direction of the polarisation in the polar coordinates (polar angle $\theta_P$ and azimuthal angle $\phi_P$). Therefore, the three components of the polarisation vector are given by:
\begin{equation}
   \begin{split}
    &P^1 = f \sin\theta_P\cos\phi_P,\\
    &P^2 = f \sin\theta_P\sin\phi_P,\\
    &P^3 = f \cos\theta_P.
\end{split} \label{eq:polvect}
\end{equation}
When $\theta_P=0$ with non-zero fraction $f$, the orientation of the polarisation is along the momentum, and the beam is longitudinally polarised. In this case, we have $P_1 = P_2 = 0$ and the polarisation matrix is diagonal. For the case that $\theta_P=\pm\pi/2$, the off-diagonal terms of the polarisation matrix would be non-zero, and the beam is transversely polarised. For the unpolarised case, the fraction $f=0$ and the polarisation matrix is the identity matrix with factor ${1}/{2}$.

\begin{figure}[h]
    \centering
    \includegraphics[width= .4\linewidth]{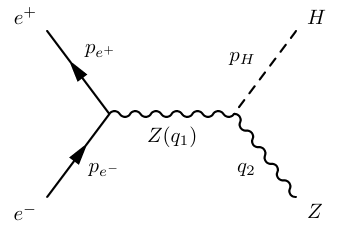}
    \caption{}
    \label{fig:diageezh}
\end{figure}

The Higgs strahlung $e^+ e^- \rightarrow ZH$ is the dominant Higgs production process at $e^+ ~e^-$ collider at $\sqrt{s}=250~$GeV, which is the main process that we focus on at the $e^+ ~e^-$ collider. The scattering amplitude of the Higgs strahlung $\mathcal{M}_{\lambda_r \lambda_u}^i$ can be easily obtained from the diagram of Fig.~\ref{fig:diageezh},
where the $\lambda_r, \lambda_u$ are the spin indices of the initial electron and positron, and the index of $i$ indicate the helicity of the radiated $Z$ boson. The unpolarised cross-section can be generated by averaging over all the helicity states of the spinor field, which implies the summation of all possible polarisation states of the electron and positron. However, since we take the polarisation of initial beams into account, we calculate the spin density matrix by applying the Bouchiat-Michel formula \cite{Bouchiat:1958yui}:
\begin{align}
    u(p,\lambda')\Bar{u}(p,\lambda)=\frac{1}{2}\left(\delta_{\lambda\lambda'}+\gamma_5 s\!\!\!/^a \sigma^a_{\lambda\lambda'}\right)(p\!\!\!/+m),\\
    v(p,\lambda')\Bar{v}(p,\lambda)=\frac{1}{2}\left(\delta_{\lambda\lambda'}+\gamma_5 s\!\!\!/^a \sigma^a_{\lambda\lambda'}\right)(p\!\!\!/-m),
\end{align}
where $\sigma^a$ is the Pauli matrices, and the four-vector $s^a_\mu,~a=1,2,3$ are the three spin vectors, which are orthogonal to each other and to the corresponding four momentum:
\begin{equation}
    \begin{split}
        &p\cdot s^a =0,\\
        &s^a \cdot s^b =\delta^{ab},\\
        &s^a_\mu s^a_\nu = -g_{\mu\nu} +\frac{p_\mu p_\nu}{m^2}.
    \end{split}
\end{equation}
Note that $\lambda$ is the spin index of the spinor field, where the eigenvalue is $\lambda=\pm\frac{1}{2}$ for the spin-1/2 particle. When the spin indices are summed over for $\lambda=\lambda'$, the spinor fields product would be recovered to the unpolarised case. In the high energy limit, the electron mass is practically negligible. In this case, the Bouchiat-Michel formula with $m\rightarrow 0$ limit is given by:
\begin{align}
    u(p,\lambda')\Bar{u}(p,\lambda)=\frac{1}{2}\left[\left(1+2\lambda\gamma_5\right)\delta_{\lambda\lambda'}+\gamma_5 \left(s\!\!\!/^1 \sigma^1_{\lambda\lambda'}+s\!\!\!/^2 \sigma^2_{\lambda\lambda'}\right)\right]p\!\!\!/\label{eq:boumi_uu}~,\\
    v(p,\lambda')\Bar{v}(p,\lambda)=\frac{1}{2}\left[\left(1-2\lambda\gamma_5\right)\delta_{\lambda\lambda'}+\gamma_5 \left(s\!\!\!/^1 \sigma^1_{\lambda\lambda'}+s\!\!\!/^2 \sigma^2_{\lambda\lambda'}\right)\right]p\!\!\!/~.
\end{align}
Thus, the spin density matrix of the Higgs strahlung process is given by:
\begin{equation}
    \rho^{ii'}_{\lambda_r\lambda_u\lambda_r'\lambda_u'} = \mathcal{M}_{\lambda_r \lambda_u}^i {\mathcal{M}^*}_{\lambda_r' \lambda_u'}^{i'}.
\end{equation}


By summing over all the helicity states of the initial states, the scattering amplitude squared would be the trace of the spin density matrix $\rho_{\lambda_r\lambda_u\lambda_r'\lambda_u'}$ multiplied by the polarisation matrices of the two initial beams:
\begin{equation}
    {|\mathcal{M}|^2}^{i i'}_{eeZH} = \mathrm{Tr}\Big(\frac{1}{2}(\delta_{\lambda_r\lambda_r'}-P^a_-\sigma^a_{\lambda_r\lambda_r'})\frac{1}{2}(\delta_{\lambda_u\lambda_u'}-P^b_+\sigma^b_{\lambda_u\lambda_u'})\rho^{i i'}_{\lambda_r\lambda_u\lambda_r'\lambda_u'}\Big),
    \label{eq:denstpol}
\end{equation}
where the spin indices of the final state $Z$ boson $i,i'$ are still open.
Eventually, the scattering amplitude squared can be divided up in the following parts, depending on the polarisation configuration:
\begin{equation}
    {|\mathcal{M}|^2}^{i i'}_{eeZH}= (1-P^3_-P^3_+)A^{i i'} + (P^3_- - P^3_+)B^{i i'} + \sum^{1,2}_{mn}P^m_- P^n_+ C^{i i'}_{mn},
    \label{eq:matpol}
\end{equation}
where the first part $A^{ii'}$ is the unpolarised scattering matrix when the polarisation vectors are both zero. In the case that only the electron beams are longitudinally polarised, the scattering matrix would be the combination of $A^{ii'}$ and $B^{ii'}$. The last part of the Eq.~\eqref{eq:matpol} $C_{mn}^{ii'}$ indicates the transverse polarisation components of the scattering matrix.

\subsection{The $Z\rightarrow\mu^-\mu^+$ decay and the angular distribution}
\begin{figure}[]
    \centering
    \includegraphics[width=.4\linewidth]{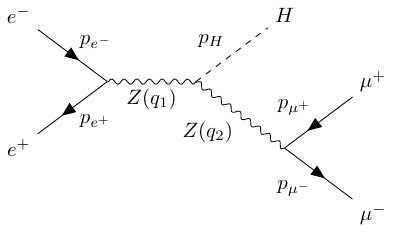}
    \caption{}
    \label{fig:eezhmumu}
\end{figure}

The polarisation of the initial beams is carried by the $Z$ boson and transferred to the final state particles by $Z$ boson decay. Since the Higgs is a scalar particle, it completely loses the spin information of the initial polarised beams. Therefore, it is more interesting to study the $Z\rightarrow \mu^- \mu^+$ decay to test the spin correlations between the initial beams and the radiated $Z$ boson, which is the process presented by the diagram of Fig.~\ref{fig:eezhmumu}

In order to take the spin correlations into account, the $Z$ decay process can be calculated by the spin density matrix $\rho^{ii'}_{Z\rightarrow \mu^+ \mu^-}$, which indicates the different helicity components of the $Z$ boson. For the production process, the spin density matrix can be obtained by the Eq.~\eqref{eq:denstpol} without summing over the helicity states of the radiated $Z$ boson. The total scattering matrix of the full process can be derived in the narrow-width approximation via contracting the polarisation states of the internal $Z$ boson:
\begin{equation}
    |\mathcal{M}|^2 \approx \frac{1}{m_Z\Gamma_Z}\sum_{ii'}{\mathcal{M}^2}^{ii'}(e^+ e^-\rightarrow Z H)\rho^{i'i}(Z\rightarrow \mu^+ \mu^-).
    \label{eq:narrowxs}
\end{equation}

Furthermore, we apply Eq.~\eqref{eq:cpmixhvv} for this process, and obtain the following form of the total amplitude squared (initial $Z$-boson polarisation already contracted):
\begin{equation}
            \begin{split}
                    |\mathcal{M}|^2 &=(1-P_-^3P_+^3)( \cos^2\xi_{\cp}\,\mathcal{A}_\text{CP-even} + {\sin2\xi_{\cp}\,\mathcal{A}_\text{CP-odd}} + \sin^2\xi_{\cp}\,\tilde{\mathcal{A}}_\text{CP-even})\\
                    &+(P_-^3-P_+^3)(\cos^2\xi_{\cp}\,\mathcal{B}_\text{CP-even} + {\sin2\xi_{\cp}\,\mathcal{B}_\text{CP-odd}} + \sin^2\xi_{\cp}\,\tilde{\mathcal{B}}_\text{CP-even})\\
                    &+\sum_{mn}^{1,2}P_-^m P_+^n\left(\cos^2\xi_{\cp}\,{\mathcal{C}}^{mn}_\text{CP-even} + {\sin2\xi_{\cp}\,\mathcal{C}^{mn}_\text{CP-odd}} + \sin^2\xi_{\cp}\,\tilde{\mathcal{C}}^{mn}_\text{CP-even} \right).
            \end{split}
            \label{eq:ampsqbsm}
            \end{equation}
Note that, all $\cp$-even terms, which are proportional to the $\cos^2\xi_\cp$ and $\sin^2\xi_\cp$, are $\cp$ conserving ($\mathcal{A}_\text{CP-even}$, $\tilde{\mathcal{A}}_\text{CP-even}$, $\mathcal{B}_\text{CP-even}$, $\tilde{\mathcal{B}}_\text{CP-even}$ and $\mathcal{C}_\text{CP-even}$ and $\tilde{\mathcal{C}}_\text{CP-even}$), while the mixing terms, proportional to $\sin2\xi_\cp$, violate the $\cp$ symmetry ($\mathcal{A}_\text{CP-odd}$, $\mathcal{B}_\text{CP-odd}$ and $\mathcal{C}_\text{CP-odd}$). The explicit analytical results of the $|\mathcal{M}|^2$ of the $e^-e ^+ \rightarrow HZ \rightarrow H \mu^- \mu^+$ process with initial beam polarisation for both the SM $\cp$-conserving cases and the BSM $\cp$-violating cases are shown in the appendix~\ref{sec:xseehmumu}. According to the analytical calculation, we know that the $\cp$-mixing terms for both unpolarised and the longitudinally polarised 
cases depend on the following triple-product:
\begin{equation}
    \mathcal{A}_\text{CP-odd}, ~\mathcal{B}_\text{CP-odd}\propto \epsilon_{\mu\nu\alpha\beta} [p_{e^-}^\mu p^\nu_{e^+} p^\alpha_{\mu^+} p^\beta_{\mu^-}]\propto \vec{p}_{e^-}\cdot (\vec{p}_{\mu^+}\times \vec{p}_{\mu^-}),
    \label{eq:trip_ul}
\end{equation}
which is related to the azimuthal-angle difference between the $e^+ e^-$ plane and the $\mu^+\mu^-$ plane in the Higgs rest frame. In the center-of-mass frame, this observable is the azimuthal-angle difference between the $ZH$ plane and the $\mu^+\mu^-$ plane.

On the other hand, the transversely polarised terms $ \mathcal{C}^{mn}_\text{CP-odd}$ can be extracted by another triple-product, which is introduced in \cite{Rao:2007ce} and given by
\begin{equation}
    \mathcal{C}^{mn}_\text{CP-odd} \propto \epsilon_{\mu\nu\rho\sigma}[({p}_{e^-} +{p}_{e^+})^\mu p_{\mu^+}^\nu p_{\mu^-}^\rho s_{e^-}^\sigma ]\propto(\vec{p}_{\mu^+}\times\vec{p}_{\mu^-})\cdot \vec{s}_{e^-}.
    \label{eq:triprd_trans}
\end{equation}
As we see in Eq.~\eqref{eq:triprd_trans}, this triple-product is the azimuthal-angle difference between the $\mu^+\mu^-$ plane and the polarisation direction of the initial beams. In this case, we define the orientation of the azimuthal plane by fixing the direction of the transverse polarisation of the electron, and the $ \mathcal{C}^{mn}_\text{CP-odd}$ directly depends on the azimuthal angle of the final state $\mu^-$.
\begin{figure}[h]
    \centering
    \includegraphics[width= .5\linewidth]{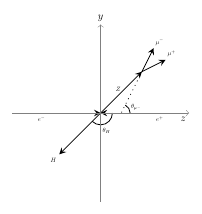}\includegraphics[width= .5\linewidth]{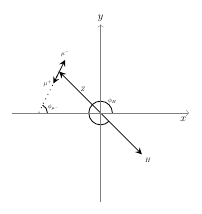}
    \caption{The coordinate system in the center of mass frame for the $e^+ e^- \rightarrow H \mu^+ \mu^-$ process, the left plot is the $y-z$ plane and the right plot is the $x-y$ plane. The direction of the electron beam is defined as the $z$-axis direction $\vec{n}_z=\Vec{p}_{e^-}/|\Vec{p}_{e^-}|$, while we choose the direction of the electron polarisation as the $y$-axis $\vec{n}_y=\Vec{s}_{e^-}/|\Vec{s}_{e^-}|$. Thus, the $x$-axis can be defined by the cross product $\vec{n}_x = (\Vec{s}_{e^-}\times\Vec{p}_{e^-})/|\Vec{s}_{e^-}\times\Vec{p}_{e^-}|$}
    \label{fig:eehmumu_coord}
\end{figure}
Therefore, we choose the center of mass frame and specify the orientation of the $x$-axis and $y$-axis by the spin vector of electron as shown in Fig.~\ref{fig:eehmumu_coord}. In this coordinate system, the cross section can be obtained by integrating over the polar angles and azimuthal angles of the Higgs boson and muon: 
\begin{equation}
    \sigma = \int \frac{|\mathcal{M}|^2}{4 s} dQ(\theta_H, \theta_{\mu^-},\phi_H, \phi_{\mu^-}).
    \label{eq:xstot}
\end{equation}
where the Lorentz invariant phase space $dQ(\theta_H, \theta_{\mu^-},\phi_H, \phi_{\mu^-})$ is shown in the appendix~\ref{sec:dlips}, and $s$ in the denominator is the center-of-mass energy squared. 
Since the total cross-section is a $\cp$-even observable, the $\cp$-mixing amplitudes in Eq.~\ref{eq:ampsqbsm} have no contribution to the total cross-section and
\begin{equation}
    \int \frac{|\mathcal{M}^\mathrm{CP-mix}|^2}{4 s} dQ(\theta_H, \theta_{\mu^-},\phi_H, \phi_{\mu^-}) =0.
\end{equation}
However, the total cross-section takes the following form
\begin{equation}
\begin{split}
        \sigma_\mathrm{tot}  &= |c_\mathrm{SM}|^2\sigma_\mathrm{SM} +2|c_\mathrm{SM}c_{HZZ}| \sigma_\mathrm{interfer} + |c_{HZZ}|^2\sigma_{HZZ} +|\widetilde{c}_{HZZ}|^2 \widetilde{\sigma}_{HZZ},
\end{split}
\label{eq:xstot}
\end{equation}
where the $\cp$-odd amplitude $c_\mathrm{SM}\widetilde{c}_{HZZ}\mathcal{M}_\mathrm{CP-vio}$ would be averaged out when we integrate over the full phase space, and this term does not contribute to the total cross-section. 
Although the total cross-section can have the contributions from $\widetilde{c}_{HZZ}$, the $\widetilde{\sigma}_{HZZ}$ is still $\cp$-even. Therefore, in order to construct a $\cp$-sensitive observable, one has to investigate the differential cross-section, particularly with respect to the azimuthal angle of final state muons.

However, the differential cross-section w.r.t the azimuthal angle would be constantly distributed, when the initial beams are unpolarised and the spin dependence would be averaged out.
In order to obtain the non-trivial azimuthal distribution, the transverse polarisation must be imposed for the initial electrons-positrons beams. 
Although the transversely polarisation yields the non-trivial distribution w.r.t the azimuthal angles, the transverse polarised amplitude $C_{mn}$ would still not contribute to the total cross-section\cite{Diehl:2003qz,Fleischer:1993ix}, because the specified azimuthal orientation would be integrated out. Therefore, only the azimuthal angular distribution would be the distinctive channel to test the $\cp$-violation effect, when we apply the transverse polarisation for the initial beams.

\section{Phenomenological analysis for the $\cp$-odd observables}
\label{sec:cpobs}

\subsection{Strategical procedure for the analysis with transversely polarised beams}
In principle, the $HZZ$ interaction is the linear combination of all the possible terms in Eq.~\eqref{eq:xstot}. However, in order to simplify the study, we can neglect the dimension 6 $\cp$-even operator with $c_{HZZ}$ in Eq.~\eqref{eq:effL}\eqref{eq:dim6hzz} and only take the $\cp$-odd term $\widetilde{c}_{HZZ}$ and the tree level SM term $c_\mathrm{SM}$ into account, since the $c_{HZZ}$ term is suppressed and does not contribute to the $\cp$-odd observables. Therefore, we set up a strategical scenario, which is assuming that the total cross-section of $e^+ e^-\rightarrow \mu^+ \mu^- H$ is only composed by the $c_\text{SM}$ and $\widetilde{c}_{HZZ}$ terms, and shown as the following
\begin{equation}
    \sigma_\text{tot} \approx |c_\text{SM}|^2 \sigma_\text{SM} + |\widetilde{c}_{HZZ}|^2 \widetilde{\sigma}_{HZZ} = \left(\cos^2\xi_{\cp}\,\kappa_\text{SM}^2  \sigma_\text{SM} + \sin^2\xi_{\cp}\frac{\widetilde{\kappa}_{HZZ}^2}{16 }\, \tilde{\sigma}_{HZZ} \right),
    \label{eq:sigtotapprox}
\end{equation}
where the cross section $\sigma_\text{SM}$ denotes the cross section in the SM at tree level, and $\widetilde{\sigma}_{HZZ}$ provides the cross-section exclusively contributed by $\widetilde{c}_{HZZ}$. In this case, the $\cp$-violation is parameterised by the $\cp$-mixing angle $\xi_{\cp}$.

 In order to explore the $\cp$-mixing impact without changing the total cross-section, we can set up a strategical scenario that the total cross-section is fixed to the tree-level SM cross-section, which means $\sigma_\text{tot} = \sigma_\text{SM}$ with $\kappa_\mathrm{SM}=1$. In this case, we can derive the condition 
 \begin{equation}
     {\widetilde{\kappa}_{HZZ}}=4\sqrt{\frac{\sigma_\text{SM}}{\widetilde{\sigma}_{HZZ}}}\sim 5.64,~\kappa_\mathrm{SM}=1.
     \label{eq:kazzfix}
 \end{equation}
Hence, the total cross-section is fixed, but the $\cp$-violation effect on the differential cross-section only depends on the $\cp$ mixing angle $\xi_{\cp}$. This scenario is helpful to test the phenomenological effect of the $\cp$-violation and to compare the exclusive $\cp$-violating result with the SM result for this specific process.

Furthermore, we can make the assumption that both initial beams are 100\% transversely polarised, and we choose the conventions that the polarisation of the electron and positron are parallel (${\phi_P}_{-}={\phi_P}_{+}=0$) and anti-parallel ${\phi_P}_{-}=0, {\phi_P}_{+}=\pi$) (see Eqs.~\eqref{eq:polvect}). One should note that, the effect of transverse polarisation can disappear when both beams are perpendicularly polarised. According to the coordinate system in Fig.\ref{fig:eehmumu_coord}, the transverse polarisation configuration for electron beams are set to along the $y$-axis, which are
\begin{align}
    &\text{parallel }&&\boldsymbol{P}_{-} = (0, 100\%,0)=\boldsymbol{P}_{+},\\
    &\text{anti-parallel }&&\boldsymbol{P}_{-} = (0, 100\%,0), ~\boldsymbol{P}_{+} = (0, -100\%,0).
\end{align}
\begin{figure}[h]
    \centering
    \includegraphics[width=.475\linewidth]{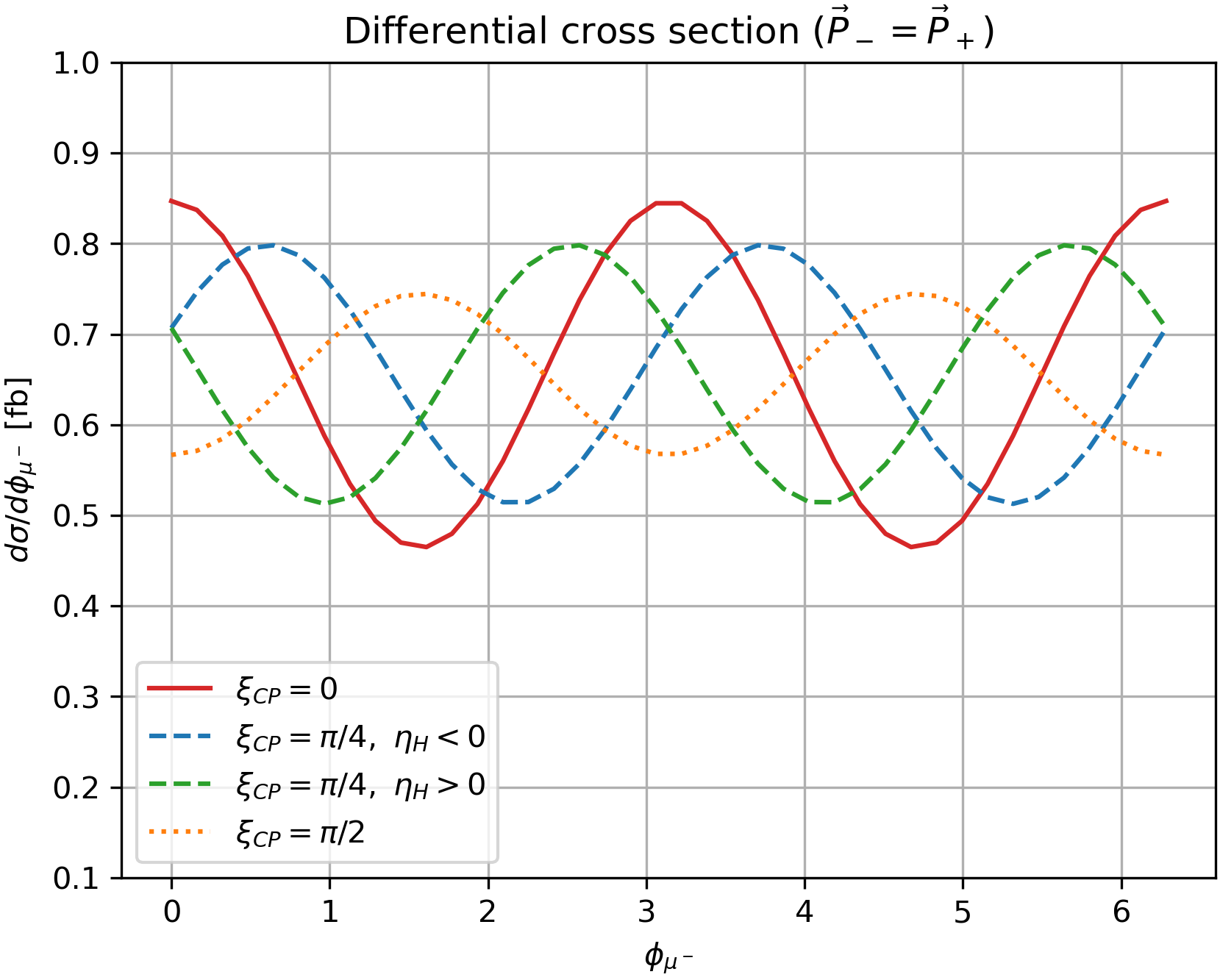}~~\includegraphics[width=.475\linewidth]{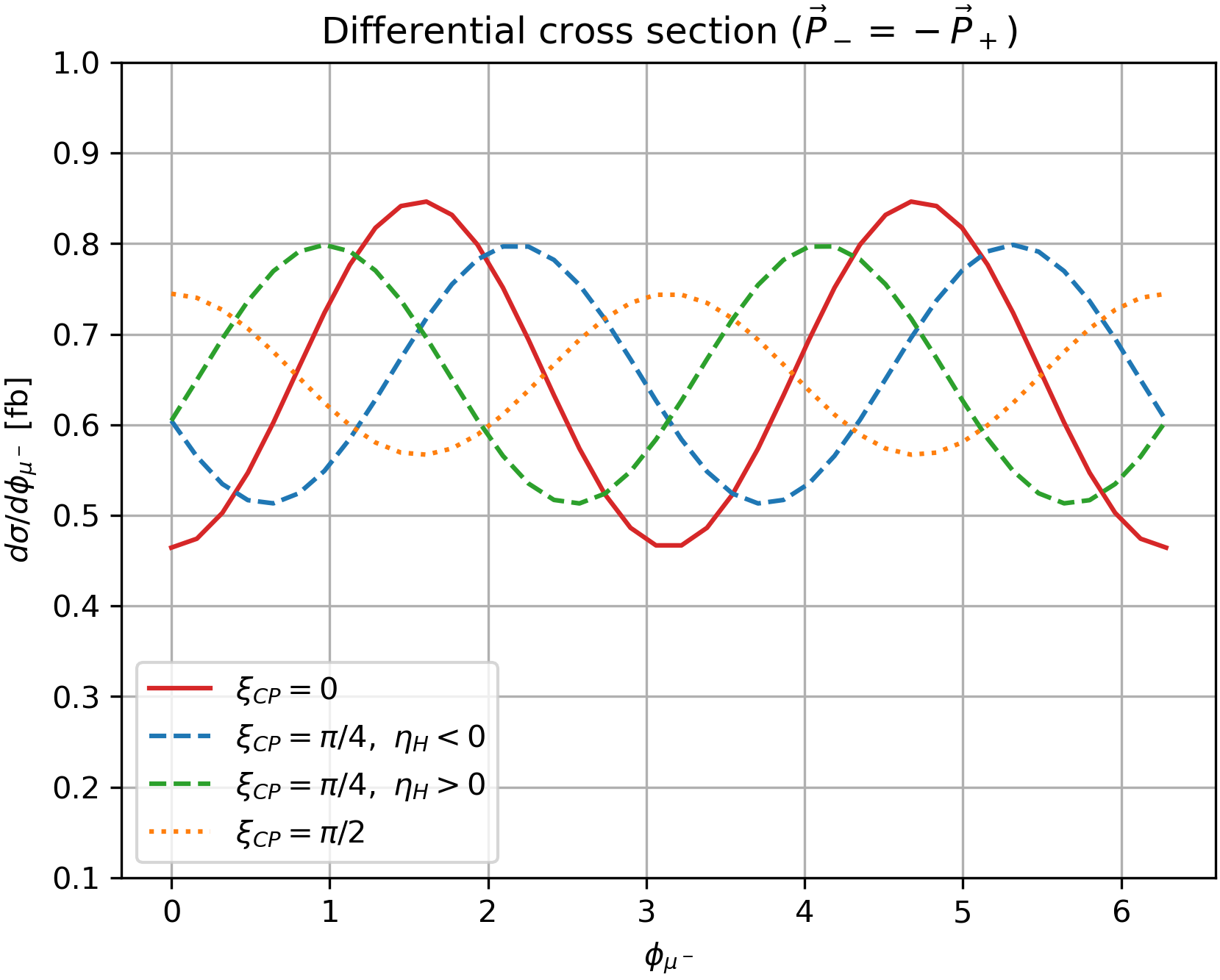}
    \caption{The analytical results of the differential cross sections with respect to the muon azimuthal angle, where the red solid lines correspond to the pure SM $\cp$-even case. The orange dotted lines demonstrate the case with only $\widetilde{\sigma}_{HZZ}$ cross section. The blue and green dashed lines are both for the $\cp$ mixing case with the mixing angle $\xi_{\cp}=\pi/4$, and correspond to forward Higgs $\eta_H>0$ and backward Higgs $\eta_H<0$ respectively (see definition in the text). The center-of-mass energy is 250 GeV. The transversely polarised beams in the left panel are parallel, and in the right panel beams are anti-parallel.}
    \label{fig:dsigmar}
\end{figure}
\begin{figure}[h]
    \centering
    \includegraphics[width=.618\linewidth]{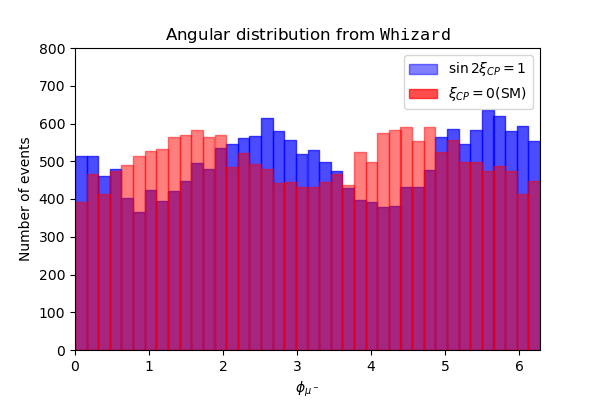}
    \caption{The Monte-Carlo simulation results of the muon azimuthal angular distribution, where the colors correspond to the same configuration as in the right panle of Fig.\ref{fig:dsigmar}. The Monte-Carlo simulation is generated by \texttt{Whizard-3.0.3}, with integrated a luminosity of 5 ab$^{-1}$ and the center-of-mass energy 250 GeV.}
    \label{fig:dndphi}
\end{figure}

In Figs.~\ref{fig:dsigmar}, we present the azimuthal angular distribution in such a strategical scenario, where the left and right panel correspond to parallel and anti-parallel polarisation configuration, respectively. In particular, the $\cp$-mixing cases with the maximal $\cp$-mixing effect $\sin2\xi_{\cp} = 1$ are separated into forward Higgs (the pseudorapidity of Higgs $\eta_H>0$ and $\cos\theta_H > 0)$ and backward Higgs (the pseudorapidity of Higgs $\eta_H<0$ and $\cos\theta_H<0)$, while the $\cp$-conserving cases lead to the same distribution for forward Higgs and backward Higgs. One can notice that the direction of the polarisation change the angular distribution, and the parallel and anti-parallel polarisation lead to the maximal effect. The non-trivial azimuthal angular distribution would vanish, when the electron and positron beams are perpendicularly polarised. 
In addition, we perform the Monte-Carlo simulation for the same strategical scenario, and show the forward Higgs $\eta_H>0 (\cos\theta_H > 0)$ with anti-parallel polarisation $(\boldsymbol{P}_{-} = - \boldsymbol{P}_{+})$ in Fig.~\ref{fig:dndphi}.


As we see, the azimuthal distribution based on the Monte-Carlo simulation basically match to the analytical result of the differential cross section, where the SM distribution of the muon azimuthal angle is symmetric under the parity transformation, i.e. is CP-even. On the other hand, the $\cp$-mixing case shifts the angular distribution to an asymmetric distribution, while the forward Higgs and backward Higgs are shifting the distribution into the opposite direction. Since the direction of the electron $e^-$ beams defines the $z$-axis of the coordinate system, the charge conjugation would flip the direction of the electron beam, and the $z$-axis would be flipped as well. However, the direction of Higgs is invariant under $\mathcal{C}$ transformation. In this case, the backward Higgs case would be changed to the forward Higgs case by the charge conjugation. Note that, the $\phi_{\mu^-}$ angular distribution of the $\cp$ mixing case can still be a constant distribution when the cases of the forward Higgs and backward Higgs are summed up together.

Based on the analysis for the angular distribution, we can construct an observable as
\begin{equation}
    \mathcal{O}^T_\cp = \eta_{H} \sin 2 \phi_{\mu^-},
    \label{eq:obscpo}
\end{equation}
which is consistent with the vector product form in \cite{Biswal:2009ar},
\begin{equation}
    \mathcal{O}^T_\cp\propto \left[\Vec{s}_{e^-}\cdot (\Vec{p}_{\mu^-}-\Vec{p}_{\mu^+}) \right] \left[(\Vec{s}_{e^-}\times \Vec{p}_{e^-})\cdot (\Vec{p}_{\mu^-}-\Vec{p}_{\mu^+}) \right][\Vec{p}_{e^-}\cdot\Vec{p}_H].
\end{equation}
In this case, the $\cp$-violation in the $HZZ$ interaction leads to differential cross-sections where the signal regions have different sign of this observable $\mathcal{O}_\cp$.

\subsection{The $\cp$-odd observable with transverse polarisation}

In order to probe the $\cp$-violation effect, one has to construct a $\cp$-odd observable. However, it is difficult to construct the actual $\mathcal{T}$-odd observable in collider experiments, since the true ``time reversal" is difficult. Consequently, we can apply the naive $\mathcal{T}$ reversal $\mathcal{T}_N$, which is the $\mathcal{T}$ reversal when neglecting all the initial and final state radiation.
If we assume that $\mathcal{CPT}\approx \mathcal{CPT}_N$, a $\mathcal{T}_N$-odd observable can be converted to a $\cp$-odd observable by the $\mathcal{CPT}$ theorem. 
Consequently, we construct an asymmetry based on the observable in Eq.~\eqref{eq:obscpo}, which is given by:
\begin{equation}
        \mathcal{A}^T_\cp = \frac{1}{\sigma_\text{tot}}\int \operatorname{sgn}(\mathcal{O}^T_\cp){d \sigma} = \frac{1}{\sigma_\text{tot}}\int d\eta_H d\phi_{\mu^-}\left( \operatorname{sgn}(\eta_H \sin2\phi_{\mu^-}) \frac{d^2\sigma}{d\eta_H d\phi_{\mu^-}}\right).
\end{equation}
In the experiment, such an asymmetry is obtained by counting the numbers of events for the two different signal regions, which is:
\begin{equation}
    \mathcal{A}^T_\cp=\frac{N(\mathcal{O}^T_\cp<0)-N(\mathcal{O}^T_\cp>0)}{N(\mathcal{O}^T_\cp<0)+N(\mathcal{O}^T_\cp>0)},
    \label{eq:asyN}
\end{equation}
where $N$ denotes the corresponding number of events. 
Since the SM is $\cp$ conserving for the neutral current, the SM background for this asymmetry is negligible. However, the number of events fluctuates statistically leading to the uncertainty of this asymmetry. The numbers of events of each region follows a Poisson distribution, which yields the statistical uncertainties $\sqrt{N}$. The uncertainty of the asymmetry, based on binomial distribution, is given by:
\begin{equation}
    \Delta {\mathcal{A}^T_\cp}=2\sqrt{\frac{\epsilon(1-\epsilon)}{N_\text{tot}}},~ \epsilon = \frac{N(\mathcal{O}_\cp^T<0)}{N_\text{tot}}.
    \label{eq:delasy}
\end{equation}
Hence, we can obtain:
\begin{equation}
    \Delta\mathcal{A}^T_\cp = \sqrt{\frac{1-{\mathcal{A}^T_\cp}^2}{N_\mathrm{tot}}}.
    \label{eq:delA}
\end{equation}
By taking the uncertainties of the asymmetry into account, one can potentially distinguish the $\CP$-mixing cases from the SM case with a given integrated luminosity and derive the unique $\cp$-violation effect.

\begin{figure}[h]
    \centering
    \includegraphics[width=.55\linewidth]{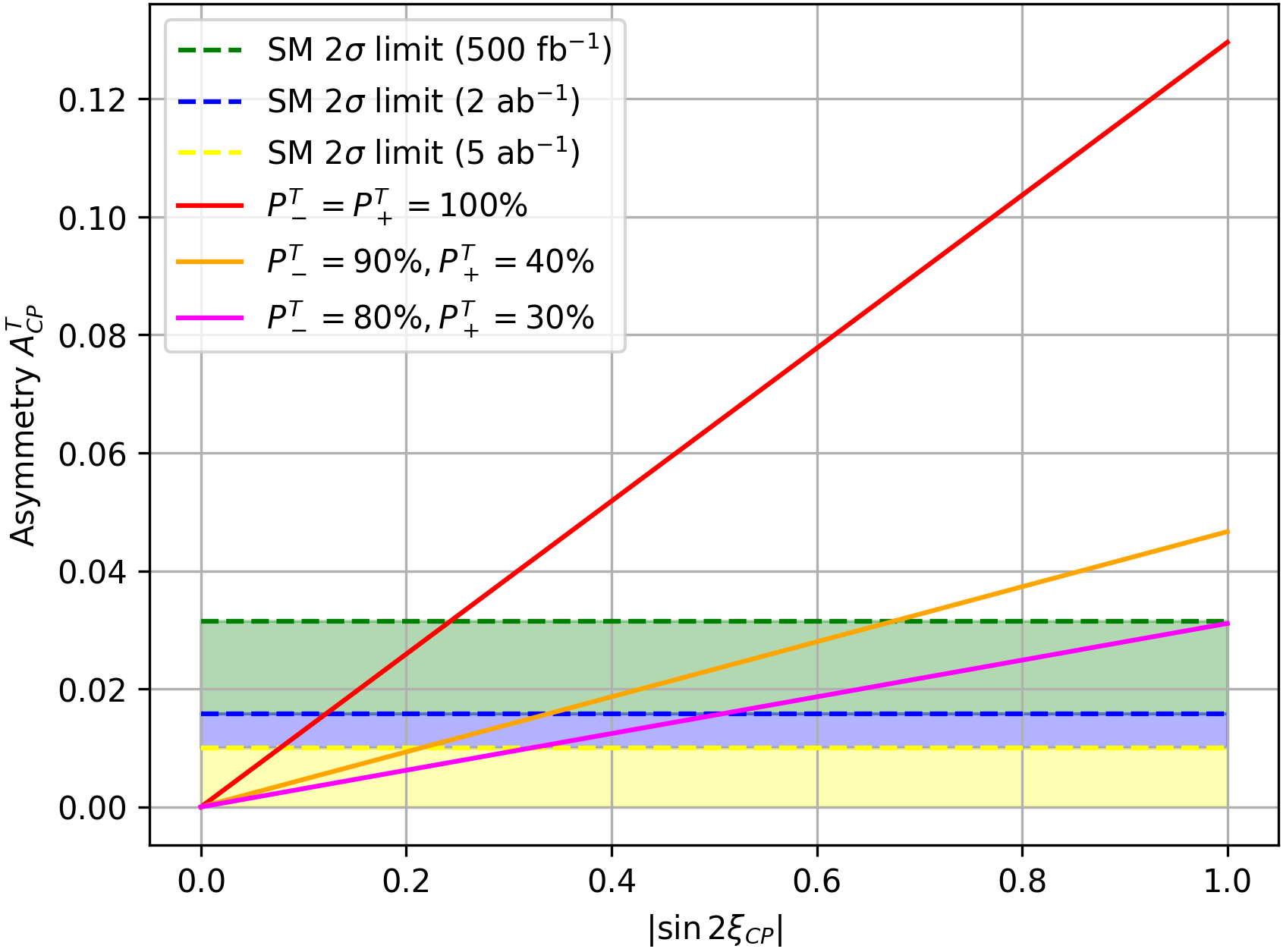}
    \caption{The analytical results of the asymmetries from Eq.~\eqref{eq:asyN} with varying $|\sin2\xi_{\cp}|$ and fixed total cross section $\sigma_\mathrm{tot}=|c_\mathrm{SM}|^2\sigma_\mathrm{SM}$, where the uncertainties of the asymmetries are taken from the Eq.~\eqref{eq:delasy}. The red solid line corresponds to the completely polarised beams $(P^T_-, P^T_+) = (100\%, 100\%)$, while the orange line and magenta line demonstrate the asymmetries with $(P^T_-, P^T_+) = (90\%, 40\%)$ and $(P^T_-, P^T_+) = (80\%, 30\%)$ polarised beams, respectively. The blue and green dashed line indicate the 2$\sigma$ limits of the asymmetry for the SM $\cp$-conserving case, while the green region is the 2$\sigma$ region of 500~fb$^{-1}$, the blue region is for the $2~\mathrm{ab}^{-1}$ and yellow region is for the $5~\mathrm{ab}^{-1}$. }
    \label{fig:asymmetry}
\end{figure}

Thus, we vary the $\CP$-mixing angles from the $
\cp$-conserving case $|\sin2\xi_{\cp}| = 0$ to the maximal $\cp$-mixing case $|\sin2\xi_{\cp}|  = 1$, and present the results of asymmetries in Fig.~\ref{fig:asymmetry}, where we still fix the total cross section $\sigma_\mathrm{tot}=\sigma_\mathrm{SM}$ as used before. 

As we see in the figure, the $\cp$-conserving case with $|\sin2\xi_{\cp}| = 0$ shows the vanishing asymmetry ${\mathcal{A}^T_\cp}$, while the $\cp$-sensitive asymmetry is enhanced with increasing $|\sin2\xi_{\cp}|$. By comparing with the SM results and its 2$\sigma$-region in Fig.~\ref{fig:asymmetry}, the $(P^T_-, P^T_+) = (80\%, 30\%)$ transversely-polarised beams cannot generate a large enough asymmetry ${\mathcal{A}^T_\cp}$, since even the ${\mathcal{A}^T_\cp}(\sin2\xi_{\cp}=1)$ is still within the 2$\sigma$ range at 500~fb$^{-1}$. However, if the integral luminosity can be increased to 2000~fb$^{-1}$, the asymmetries for $|\sin2\xi_{\cp}|\gtrsim0.5$ are above the blue region, which can be roughly distinguished from the SM $\cp$-conserving case at 95\% C.L. (Confidence Level). Furthermore, we can use the $(P^T_-, P^T_+) = (90\%, 40\%)$ transversely-polarised beams, which are the maximum polarisation fraction for the electron and positron beams expected to be obtained by experiment \cite{Moortgat-Pick:2005jsx}. In this case, the limit of $|\sin2\xi_{\cp}|$, where the asymmetry ${\mathcal{A}^T_\cp}$ can be distinguished from the SM $\cp$-conserving case, can be improved by the increment of the polarisation fraction.

\begin{figure}[h]
    \centering
    \includegraphics[width=.55\linewidth]{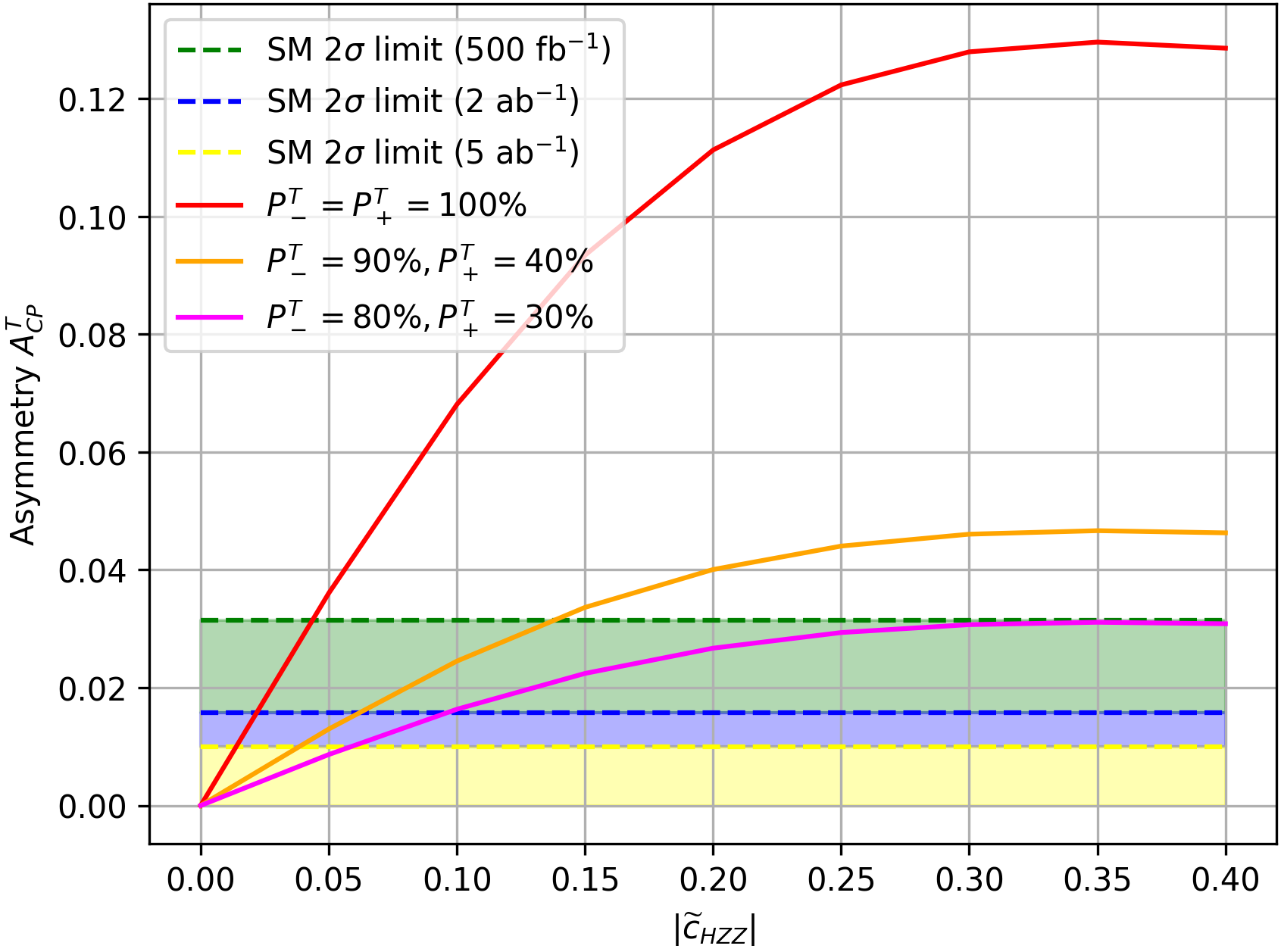}
    \caption{The analytical results of the asymmetries from Eq.~\eqref{eq:asyN} depending on the coupling $\widetilde{c}_{HZZ}$, where the SM tree-level cross-section is fixed with the additionally varying $\widetilde{c}_{HZZ}$.The configurations of polarisation and luminosity as well as the uncertainties are presented with the same colors as in Fig.~\ref{fig:asymmetry}}
    \label{fig:asymmetry2}
\end{figure}

The actual total cross section would be the linear combinations of all three possible terms in Eq.~\eqref{eq:effL}, where the size of each contributions remains unknown. Thus, this observable can be also used for a complementary measurement of the $\cp$-odd coupling $\widetilde{c}_{HZZ}$, when the $\cp$-odd coupling contribute the total cross-section. In such case, we can fix the SM tree-level contribution by $c_\mathrm{SM}=1$ and vary the $\widetilde{c}_{HZZ}$ individually, while the results of asymmetry ${\mathcal{A}^T_\cp}$ in such scenario would be presented in Fig.~\ref{fig:asymmetry2}. This figure demonstrates that the asymmetry ${\mathcal{A}^T_\cp}$ reaches the maximum when $\widetilde{c}_{HZZ}\approx 0.4$, where the $\cp$-odd and $\cp$-even interaction contribute the same amount to the total cross-section. 


Note that, the maximum values of ${\mathcal{A}^T_\cp}$ can be suppressed by smaller transverse polarisation fraction, where the $(P^T_-, P^T_+) = (80\%, 30\%)$ polarised beams with the luminosity of 500~fb$^{-1}$ cannot generate a large enough maximum asymmetries beyond the SM 2$\sigma$ deviation. Hence, the $(P^T_-, P^T_+) = (80\%, 30\%)$ polarisation with 500~fb$^{-1}$ is insufficient to determine the $\cp$ structure of $HZZ$ interaction in any cases. However, the luminosity of 2~ab$^{-1}$ can improve this sensitivity significantly, and the fraction $\widetilde{c}_{HZZ}\sim 0.1$ can be determined by $(P^T_-, P^T_+) = (80\%, 30\%)$ using asymmetry ${\mathcal{A}^T_\cp}$.

 \subsection{The $\cp$-odd observable with unpolarised or longitudinally polarised beams}
 \label{sec:obslong}
  \begin{figure}[h]
     \centering
     \includegraphics[width=.618\linewidth]{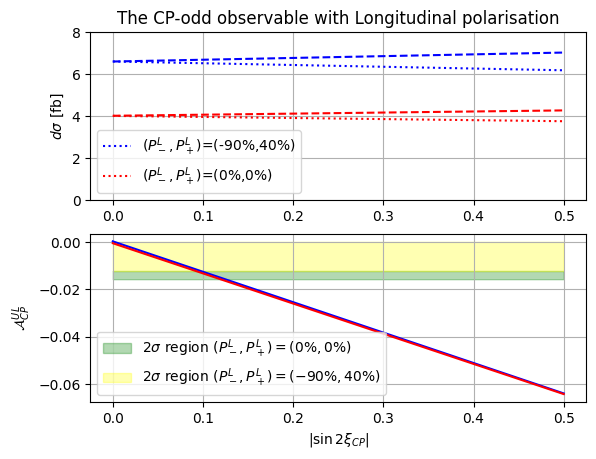}
     \caption{Plots of the asymmetry $\mathcal{A}_\cp^{UL}$ (see Eq.~\eqref{eq:aul}) and the partial cross-section with varying $\sin2\xi_{\cp}$ at the ILC with the center-of-mass energy $\sqrt{s}=250$~GeV. The red lines in both panels are for the unpolarised case, and the blue lines correspond to the case with longitudinally polarised beams $(P^L_-, P^L_+)=(-90\%, 40\%)$. The upper panel illustrates the cross-section with different signs of the observable $\mathcal{O}^{UL}_\cp$, where the dotted lines are for the cross-section $\sigma(\mathcal{O}^{UL}_\cp<0)$ and the dashed lines are for the $\sigma(\mathcal{O}^{UL}_\cp>0)$. The lower plot presents the asymmetry $\mathcal{A}_\cp^{UL}$ for both unpolarised and longitudinally polarised cases. The yellow region is the $2\sigma$ region of SM $\cp$-conserving case with $(P^L_-, P^L_+)=(-90\%, 40\%)$ and 2~ab$^{-1}$, while the green region is for the unpolarised case.}
     \label{fig:asy_longp}
 \end{figure}
 In addition to the observable in Eq.~\eqref{eq:obscpo}, there is another observable, which is sensitive to the triple product in the unpolarised and longitudinally polarised cross-section in Eq.~\eqref{eq:trip_ul}, and shown in the following,
 \begin{equation}
     \mathcal{O}^{UL}_\cp = \cos\theta_{\mu}\sin\Delta\phi_{H\mu}\propto (\Vec{p}_{\mu^-}\times \Vec{p}_{\mu^+})\cdot\Vec{p}_H, \qquad \Delta\phi_{H\mu} = \phi_\mu - \phi_H.
     \label{eq:obsul}
 \end{equation}
 This observable can be measured for any kind of initial beams polarisation, and can be used to construct another asymmetry 
 \begin{equation}
     \mathcal{A}_\cp^{UL} = \frac{1}{\sigma_\mathrm{tot}}\int\mathrm{sgn}(\mathcal{O}_\cp^{UL})d\sigma = \frac{N(\mathcal{O}^{UL}_\cp < 0) - N (\mathcal{O}^{UL}_\cp >0)}{N(\mathcal{O}^{UL}_\cp < 0) + N (\mathcal{O}^{UL}_\cp >0)},
     \label{eq:aul}
 \end{equation}
where the statistical uncertainty of this asymmetry can be obtained by the same formula, see Eq.~\eqref{eq:delA}.

We calculate this asymmetry and differential cross-section w.r.t the $\cp$-odd observable $\mathcal{O}^{UL}_\cp$ in Fig.\ref{fig:asy_longp}, where the upper panel shows that longitudinal polarisation can enhance the total cross-section of the process $e^+ e^-\rightarrow H \mu^+ \mu^-$. As we see in the lower panel of Fig.\ref{fig:asy_longp}, the asymmetries $\mathcal{A}_\cp^{UL}$ of both the unpolarised case and the longitudinal polarisation are basically the same, which are also linearly depend on the $\cp$-mixing angle $\sin2\xi_\cp$. However, due to the larger total cross-section, the statistical uncertainty can be suppressed and the precision of measuring the $\mathcal{A}_\cp^{UL}$ is getting better when the longitudinal polarisation is imposed. For the integrated luminosity of 2~ab$^{-1}$, however, the $\mathcal{A}_\cp^{UL}$ with the $|\sin2\xi_\cp|\gtrsim 0.1$ can be 2$\sigma$ different from the $\cp$-conserving value. This sensitivity to the $\cp$-violation effect is better than the measurement of $\mathcal{A}^T_\cp$ with the transverse polarisation and the same luminosity, shown in Fig.~\ref{fig:asymmetry}. This is due to the suppression of the observable $\mathcal{A}^T_\cp$ by the 
prefactor of polarisation degree $P^T_- P^T_+ < 1$, while $\mathcal{A}_\cp^{UL}$ is originating from the unpolarised part and has therefore no suppression from polarisation degrees. 

However, the observable $\mathcal{A}_\cp^{UL}$ can be also measured when the initial beams are transversely polarised, since the triple product in Eq.~\eqref{eq:obsul} exists in the unpolarised cross-section and can still contribute in such a case. Therefore, one can measure two $\cp$-odd observables simultaneously with imposing transverse polarisation, which are $\mathcal{A}^{UL}_\cp$ and $\mathcal{A}^T_\cp$. In such a case, the sensitivity to the $\cp$-odd effect can be improved furthermore by combining these two observable measurements.

\section{The determination limits of the $\cp$-violation with beam polarisation at the ILC}
In the previous section, we discuss the impact of the $\cp$ sensitive observables. Hence, we can use these observables to determine the size of the $\cp$-violation effect at the ILC with certain integrated luminosities and polarisation degrees, where we used two scenarios for the determinations. One of the scenario is i) {fixing the total cross-section}, while only the $\cp$ property of the process can be varied. In such a case, one can determine the intrinsic $\cp$-mixing angle with the help of asymmetries in section \ref{sec:cpobs}. The another scenario is supposing that the ii) {SM tree-level cross-section is fixed}, and the $\widetilde{c}_{HZZ}$ term contribute the total cross-section additionally. In this case, the total cross-section can be varied by the $\cp$-odd coupling. Therefore, we can determine the $\cp$-odd coupling by fitting the numbers of events in the signal regions differed by $\cp$-odd observables.

\subsection{The determination for the $\cp$-mixing angle}

\begin{figure}[h]
    \centering
    \includegraphics[width=.475\linewidth]{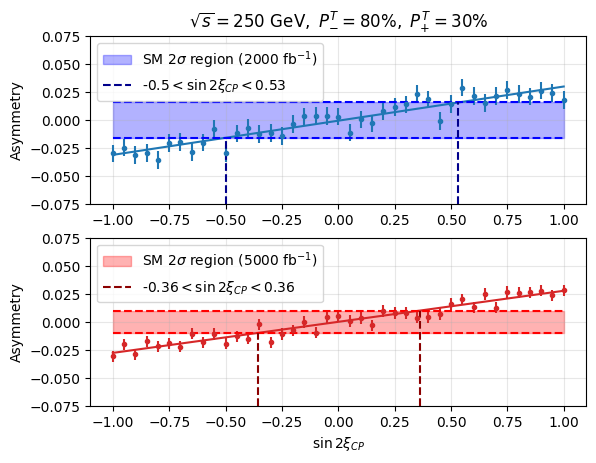}~~\includegraphics[width=.475\linewidth]{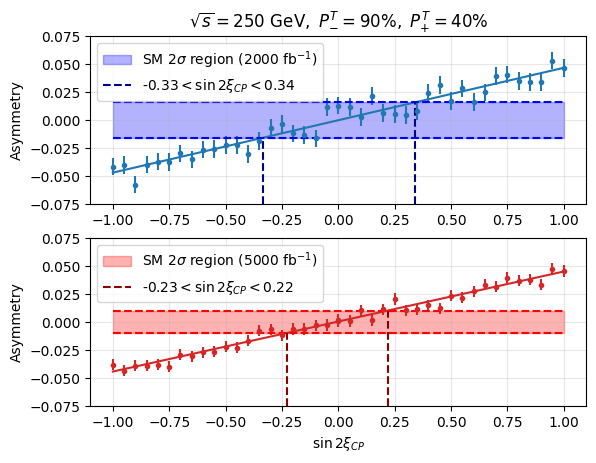}
    \caption{The plots of asymmetry $\mathcal{A}^T_\cp$ (see Eq.~\eqref{eq:asyN}) vs $\sin2\xi_\cp$. The dotted points with error bars are the Monte-Carlo simulation results of the $\mathcal{A}^T_\cp$ generated by the $\texttt{Whizard-3.0.3}$, where the error bars are the statistical uncertainties and the total cross-section is fixed to the SM tree-level value. The solid lines are the linear fit of the MC data. The left panels correspond to the polarisation fraction $(P^T_-, P^T_+) = (80\%, 30\%)$, and the right panels correspond to $(P^T_-, P^T_+) = (90\%, 40\%)$. The SM 2$\sigma$-bounds of the upper two plots refer to the integrated luminosity of 2000~fb$^{-1}$, and the lower two plots refer to 5000~fb$^{-1}$. For all the figures, the fitting lines come across the SM 2$\sigma$-bounds determining the limit of $\sin2\xi_\cp$.}
    \label{fig:cpmixingscan1}
\end{figure}

We present the Monte-Carlo simulation results of the asymmetry $\mathcal{A}^T_\cp$ with varying $\cp$-mixing angle in Figs.~\ref{fig:cpmixingscan1}, which are generated by $\mathtt{Whizard-3.0.3}$. As we see in Figs.~\ref{fig:cpmixingscan1}, the asymmetry is linear dependent on $\sin2\xi_\cp$, which is the same as the analytical calculation in Fig.~\ref{fig:asymmetry}, where the asymmetry has the bigger statistical fluctuation for the integral luminosity $2000~\mathrm{fb}^{-1}$ than for the $5000~\mathrm{fb}^{-1}$. Since the $500~\mathrm{fb}^{-1}$ is insufficient to determine the $\cp$-violation effect based on previous discussions, we do not present the results with $500~\mathrm{fb}^{-1}$ for the Monte Carlo simulation. Based on the MC data, we perform linear fits for the $\mathcal{A}^T_\cp$ - $\sin2\xi_cp$ dependence, which are presented in the solid lines in Figs.~\ref{fig:cpmixingscan1}. By comparing with the SM 2$\sigma$-limits with respect to different integrated luminosities, we obtain the limits of the $\cp$-mixing parameter $\sin2\xi_\cp$ with different transverse polarisation fractions and different asymmetries, which are presented in Tab.~\ref{tab:sum_xi}. In these studies, we do not consider the background estimation, since the SM background is basically $\cp$-even and the asymmetries would cancel out the $\cp$-even contribution. Therefore, we can simply estimate the limits of the $\cp$-mixing parameters without taking the background into account.
Since the unpolarised observable $\mathcal{A}^{UL}_\cp$ can be simultaneously measured when transverse polarisation is imposed, we can combine the two observables by introducing the following $\chi^2$: 
\begin{equation}
    \chi^2_{\mathcal{A}_\cp} = (\frac{\mathcal{A}^T_\cp}{\Delta\mathcal{A}^T_\cp})^2 + (\frac{\mathcal{A}^{UL}_\cp}{\Delta\mathcal{A}^{UL}_\cp})^2.
\end{equation}
We take the 95\% C.L. of one degree of freedom as the critical value of $\chi^2_{\mathcal{A}_\cp}$, which is roughly $\chi^2_{\mathcal{A}_\cp}<3.84$. Consequently, these determination results are shown in Tab.~\ref{tab:sum_xi} as well.

Furthermore, we can also determine the $\cp$-violation with only using the longitudinal polarisation. Although the $\cp$-odd observable $\mathcal{A}^{UL}_\cp$ cannot be enhanced by the longitudinal polarisation, the total cross-section would be enlarged and the determination results can be improved. As a result, we also present the determination results using longitudinally polarised beams in Tab.~\ref{tab:sum_xi}.

\begin{table}[h]
    \centering
                 \resizebox{\linewidth}{!}{\begin{tabular}{cc|ccc}
                 \hline
                  $(P_-, P_+)$& $\mathcal{L}$ [ab$^{-1}$]& \multicolumn{3}{c}{$\sin 2\xi_{\cp}$ limit}\\
                  \multicolumn{2}{c|}{Observables} & $\mathcal{A}^T_\cp$&  Combine $\mathcal{A}^T_\cp$ \& $\mathcal{A}^{UL}_\cp$&  $\mathcal{A}^{UL}_\cp$\\
                  \hline
                  \multicolumn{2}{c}{Transverse polarisation}&&&\\
                  \hline
                  $(80\%, 30\%)$&    2.0 &  [-0.50, 0.53]&[-0.113, 0.125]&\\
                  $(80\%, 30\%)$&    5.0 &  [-0.36, 0.36]&[-0.068, 0.079]&\\
                  $(90\%, 40\%)$&    2.0 &  [-0.33, 0.34]&[-0.118, 0.110]&\\
                  $(90\%, 40\%)$&    5.0 &  [-0.23, 0.22]&[-0.066, 0.077]&\\
                  $(100\%, 100\%)$&    5.0 &  [-0.082, 0.069]&[-0.056, 0.051]&\\
                  \hline
                  \multicolumn{2}{c}{Longitudinal polarisation}&&&\\
                  \hline                  
                  $(-80\%, 30\%)$&    2.0 &  &&[-0.119,0.082]\\
                  $(-80\%, 30\%)$&    5.0 &  &&[-0.066,0.063]\\
                  $(-90\%, 40\%)$&    2.0 &  &&[-0.085,0.106]\\
                  $(-90\%, 40\%)$&    5.0 &  &&[-0.059,0.062]\\
                  $(-100\%, 100\%)$&    5.0 &  &&[-0.047,0.053]\\
\hline
                 \end{tabular}}
    \caption{The summary table for 2$\sigma$ limit of $\cp$-mixing angle $\sin2\xi_\cp$ with center-of-mass energy 250~GeV. The column of $\mathcal{A}^T_\cp$ shows, the determination results only using the observable $\mathcal{A}^T_\cp$ with transverse polarisation, while the column of $\mathcal{A}^{UL}_\cp$ corresponds to the results with using longitudinal polarisation. Note that the column of ``Combine $\mathcal{A}^T_\cp$ \& $\mathcal{A}^{UL}_\cp$" still uses the experimental set up of transverse polarisation but measures the two observables.}
    \label{tab:sum_xi}
\end{table}

As we see in Tab.~\ref{tab:sum_xi}, the method of combining the two observable with transverse polarisation yields much better precision for the $\cp$-mixing angle $\sin2\xi_\cp$ than the method of only using $\mathcal{A}^T_\cp$. Although the longitudinal polarisation can not enhance the $\cp$-odd observable, the sensitivity to the $\cp$-violation effect can be still improved by the longitudinally polarised beams due to the larger total cross-section. Consequently, the precision of using the longitudinal polarisation can be approximately the same or even slightly better than using the transverse polarisation and combining the two observables.

\subsection{The determination for the $\cp$-odd coupling}
If we assume that the SM tree-level contribution of this process is fixed and the $\widetilde{c}_{HZZ}$ term provides an additional contribution, the total cross-section can be increased by the $\cp$-odd coupling. In order to take the effect of cross-section increment into account,
we perform the fit for the corresponding signal regions deferred by the $\cp$ sensitive observable, and obtain $\chi^2$ by
\begin{equation}
    \chi_N^2 = \sum_i\left( \frac{(N(\mathcal{O}_i<0)- N^\mathrm{SM}(\mathcal{O}_i<0))^2}{N(\mathcal{O}_i<0)} +\frac{(N(\mathcal{O}_i>0)- N^\mathrm{SM}(\mathcal{O}_i>0))^2}{N(\mathcal{O}_i>0)}\right),
    \label{eq:llh}
\end{equation}
where $i$ corresponds to the different $\cp$-violating observables.
For this analysis, we only use statistical uncertainties for the rough estimation without including the systematic uncertainties.

\begin{figure}[h]
    \centering
    \includegraphics[width=.6\linewidth]{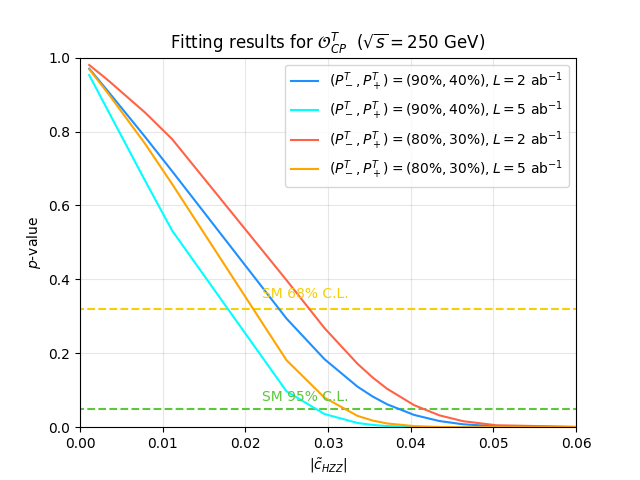}
    \caption{The $p$-value of $\chi^2_N$ function in Eq.~\eqref{eq:llh} depending on the  $\cp$-odd coupling $\widetilde{c}_{HZZ}$, where the observable is only for the $\mathcal{O}^T_\cp$. The red and orange solid lines are both using the transverse polarisation $(P^T_-, P^T_+) = (80\%, 30\%)$, and corresponds to the integrated luminosity of 2~ab$^{-1}$ and 5~ab$^{-1}$ respectively. The blue and cyan lines are using polarised beams $(P^T_-, P^T_+) = (90\%, 40\%)$ and integrated luminosity of 2~ab$^{-1}$ and 5~ab$^{-1}$ respectively. The area below the green dashed line is the region deviated from SM at 95\% C.L., while the yellow dashed line is for the SM 68\% C.L..}
    \label{fig:fcp_analy}
\end{figure}

Fig.~\ref{fig:fcp_analy} presents the $p$-value of the $\chi^2_N$ fit in Eq.~\eqref{eq:llh}, where the observable is only referring to $\mathcal{O}^T_\cp$, and Fig.~\ref{fig:fcp_analy} demonstrates the $p$-value dependence on the coupling $\widetilde{c}_{HZZ}$ obtained by analytical calculation. By comparing the solid lines with 95\% C.L., one can easily determine the limit of $\widetilde{c}_{HZZ}$, where 5~ab$^{-1}$ luminosity provides a limit of $\widetilde{c}_{HZZ}<0.03$. In particular, the higher luminosity of 5~ab$^{-1}$ with lower polarisation degrees $(P^T_-, P^T_+) = (80\%, 30\%)$ provides the better precision in $\widetilde{c}_{HZZ}$ than $(P^T_-, P^T_+) = (90\%, 40\%)$ but with lower luminosity of 2~ab$^{-1}$.

\begin{figure}[h]
    \centering
    \includegraphics[width=.5\linewidth]{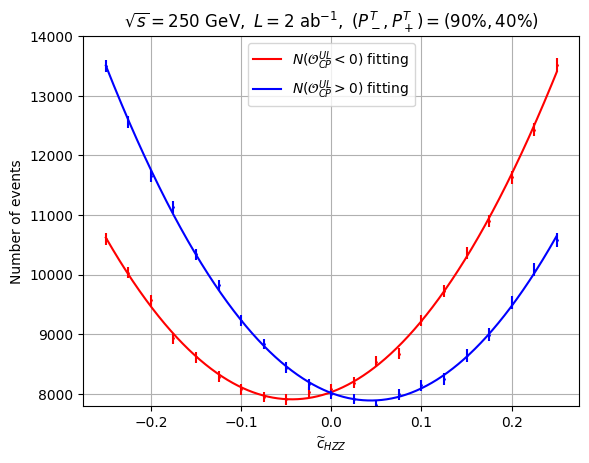}\includegraphics[width=.5\linewidth]{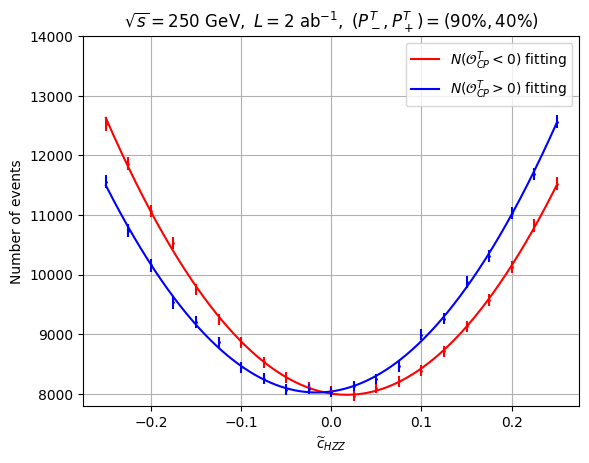}
    \caption{The quadratic fitting function result of the number of events in different signal regions with respect to the $\cp$-odd coupling $\widetilde{c}_{HZZ}$. The red lines and data points are for the signal region with $\mathcal{O}_\cp<0$, and the blue lines and data points are for $\mathcal{O}_\cp>0$. The left panel is for the signal regions defined by the signs of $\mathcal{O}^{UL}_\cp$, and the right panel corresponds to the observable $\mathcal{O}^{T}_\cp$. Both cases are using the transverse polarised beams of $(P^T_-, P^T_+) = (90\%, 40\%)$ and integrated luminosity of 2~ab$^{-1}$.}
    \label{fig:nfit}
\end{figure}

\begin{figure}[h]
    \centering
\includegraphics[width=\linewidth]{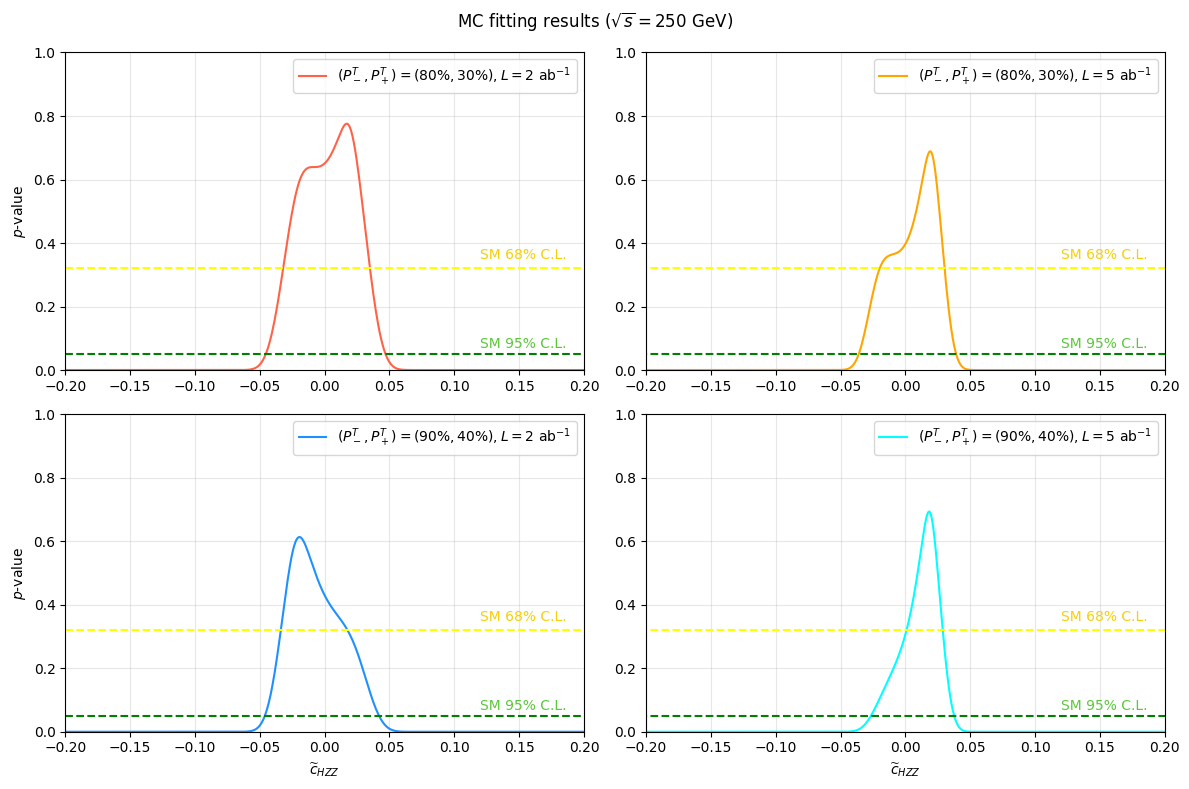}
    \caption{The $p$-values of the $\chi^2$ function defined in Eq.~\ref{eq:llh} depending on the coupling $\widetilde{c}_{HZZ}$, where the model predictions are generated by fitting the \texttt{Whizard-3.0.3} simulation data. The upper two plots are both for the polarisation $(P^T_-, P^T_+) = (80\%, 30\%)$, and the lower two plots are for $(P^T_-, P^T_+) = (90\%, 40\%)$. The left two plots use the integrated luminosity of 2~ab$^{-1}$, and the right two plots use 5~ab$^{-1}$. Both yellow and green dashed lines correspond to the SM 68\% C.L. and 95\% C.L..}
    \label{fig:fcpvcan}
\end{figure}

Furthermore, we implement the fit method for the Monte-Carlo data generated by \texttt{Whizard-3.0.3}, where we made the quadratic function fitting to the number of events in each signal regions with respect to the coupling $\widetilde{c}_{HZZ}$. The fit function is shown as the following
\begin{equation}
    N=a\,\widetilde{c}_{HZZ}^2 +b\,\widetilde{c}_{HZZ} + c.
\end{equation}
where the uncertainties of $N$ are obtained by the statistical fluctuation. Here, two of the fitting results are shown in Figs.~\ref{fig:nfit} as examples.
By using the number of events determined via the fitting lines, one can calculate the $\chi^2_N$ function in Eq.~\ref{eq:llh} and obtain the statistical $p$-values for specific polarisation fractions and luminosities, shown in Figs.~\ref{fig:fcpvcan}.

\begin{table}[h]
    \centering
    \resizebox{\linewidth}{!}{\begin{tabular}{cc|ccc}
    \hline
         $(P_{-},P_{+})$& Luminosity [ab$^{-1}$]& \multicolumn{3}{|c}{$\widetilde{c}_{HZZ} ~(\times 10^{-2})$ limit}  \\
         \multicolumn{2}{c|}{Observables} & $\mathcal{O}^{T}_\cp$& Combine $\mathcal{O}^{UL}_\cp$ \& $\mathcal{O}^T_\cp$& $\mathcal{O}^{UL}_\cp$\\
         \hline
         \multicolumn{2}{c|}{Transverse polarisation}& &&\\
         \hline
         (80\%, 30\%)&    2.0 & [-4.45,4.65]& [-2.26, 1.93]&  \\
         (80\%, 30\%)&    5.0 & [-3.55,3.85]& [-1.29, 1.06]&  \\
         (90\%, 40\%)&    2.0 & [-4.55,4.15]& [-2.24, 1.69]&  \\
         (90\%, 40\%)&    5.0 & [-2.65,3.75]& [-1.12, 0.98]&  \\
         \hline
         \multicolumn{2}{c|}{Longitudinal polarisation}& &&\\
         \hline
         $(-80\%, 30\%)$&    2.0 &  &&[-1.55,1.96]\\
                  $(-80\%, 30\%)$&    5.0 &  &&[-1.01,1.16]\\
                  $(-90\%, 40\%)$&    2.0 &  &&[-1.73,1.53]\\
                  $(-90\%, 40\%)$&    5.0 &  &&[-0.93,1.18]\\
         \hline
    \end{tabular}}
    \caption{The summary table for the limits of $\cp$-odd coupling $\widetilde{c}_{HZZ}$ at 95\% C.L., where the results with using transverse and longitudinal polarisation are both presented in the table. Particularly, the results with transverse polarisation are including the fitting only referring to $\mathcal{O}^T_\cp$ and the fitting combining $\mathcal{O}^T_\cp$ and $\mathcal{O}^{UL}_\cp$.
    The center-of-mass energy are both 250~GeV, and the polarisation fractions are using $(80\%, 30\%)$ and $(90\%, 40\%)$, while the integrated luminosities are 2~ab$^{-1}$ and 5~ab$^{-1}$.}
    \label{tab:sum_fcp}
\end{table}

Consequently, we are able to determine a limit of $\widetilde{c}_{HZZ}$ coupling by comparing the $p$-value lines with SM 95\% C.L. level in Figs.\ref{fig:fcpvcan}, and all the results with the different possible experimental configurations are presented in Tab.~\ref{tab:sum_fcp}. As we see in Tab.~\ref{tab:sum_fcp}, the determination with only using $\mathcal{O}^T_\cp$ yields a limit of $\widetilde{c}_{HZZ} \sim 0.03$, where the initial beams are $(P^T_-, P^T_+) = (90\%, 40\%)$ polarised and the integrated luminosity is 5~ab$^{-1}$. However, combining $\mathcal{O}^T_\cp$ and $\mathcal{O}^{UL}_\cp$ can strongly improve the sensitivity to $\cp$-odd coupling, and provides a limit of $\widetilde{c}_{HZZ} \sim 0.01$. Note that the higher polarisation fraction cannot significantly enhance the precision of the $\cp$-odd coupling while the integrated luminosity is fixed. Nevertheless, the limits of $\widetilde{c}_{HZZ}$ can be more precise with larger luminosity for fixed polarisation degrees. In addition, we also present the results with using longitudinal polarisation only in Tab.~\ref{tab:sum_fcp}. One can see that the $(P^L_-, P^L_+) = (-90\%, 40\%)$ polarisation and 5~ab$^{-1}$ luminosity can determine the limit of $\cp$-odd coupling $\widetilde{c}_{HZZ} \sim 0.01$, which is roughly the same as the result with transverse polarisation of $(P^T_-, P^T_+) = (90\%, 40\%)$ and $\mathcal{L}=$5~ab$^{-1}$. However, for the configuration of $(80\%, 30\%)$ and 2~ab$^{-1}$, the result with using only longitudinal polarisation gives $\widetilde{c}_{HZZ} \sim 0.017$ occasionally better than the result with only using transverse polarisation, $\widetilde{c}_{HZZ} \sim 0.02$.

In the end, we summarize the current measurements of the $HZZ$ coupling and the analyses at other future colliders in Tab.~\ref{tab:comp_exp}, where the interpretations of the other analyses can be translated by the relations given in appendix~\ref{sec:matching}. As we see, the ILC 250 GeV with transverse or longitudinal polarisations $(|P_{-}|,|P_{+}|)=(90\%,40\%)$ and 5000~fb$^{-1}$ can significantly improve the precision of the $\widetilde{c}_{HZZ}$ coupling compared to current ATLAS~\cite{ATLAS:2023mqy} and CMS~\cite{CMS:2017len,CMS:2019jdw} results.
Regarding expected HL-LHC results \cite{Cepeda:2019klc}, this method accessible at $e^+ e^-$ colliders can determine the $\widetilde{c}_{HZZ}$ coupling much better than the hadron collider with 3~ab$^{-1}$. Note that the polarised beams at $e^+e^-$ collider can improve the sensitivity to the $\cp$-odd coupling, compared to the CEPC unpolarised analysis via the exact same Higgs strahlung process with 5.6 ab~$^{-1}$ \cite{Sha:2022bkt}. However, the determination of the $\widetilde{c}_{HZZ}$ coupling via $Z$-fusion at 1~TeV CLIC \cite{Vukasinovic:2023jxd} can also provide a sensitivity to $\cp$-odd couplings roughly at the same level as the 250~GeV ILC results with polarisation. Since the $Z$-fusion process is the different channel to the Higgs strahlung process, and can be more dominant with larger center-of-mass energy, the $Z$-fusion analysis at CLIC would be the complementary study for $\cp$-violation of $HVV$ interaction.

\begin{table}[h]
    \centering
    \resizebox{\linewidth}{!}{\begin{tabular}{c|ccccccc}
    \hline
       Experiments&  ATLAS\cite{ATLAS:2023mqy}& CMS\cite{CMS:2021nnc}& HL-LHC\cite{Cepeda:2019klc}& CEPC\cite{Sha:2022bkt}& CLIC\cite{Karadeniz:2019upm}& CLIC \cite{Vukasinovic:2023jxd,BozovicJelisavcic:2024czi}& ILC\\
       Processes& $H\rightarrow 4\ell$& $H\rightarrow 4\ell$& $H\rightarrow 4\ell$& $ H Z$&  $W$-fusion& $Z$-fusion &$ H Z,~Z\rightarrow \mu^+\mu^-$\\
       $\sqrt{s}$ [GeV]&  13000&  13000& 14000& 240& 3000& 1000& 250\\
       Luminosity~[fb$^{-1}$]&  139& 137& 3000& 5600& 5000& 8000& 5000\\
       $(|P_{-}|,|P_{+}|)$& & & & & & & $(90\%, 40\%)$\\
       \hline
       $\widetilde{c}_{HZZ}~(\times 10^{-2})$ &&&&&&\\
       95\% C.L. (2$\sigma$)limit&   [-16.4, 24.0]& [-9.0, 7.0]& [-9.1, 9.1]& [-1.6, 1.6]& [-3.3, 3.3]& [-1.1, 1.1]&  [-1.1, 1.0] \\
       \hline
    \end{tabular}}
    \caption{Summary of the limits of $\widetilde{c}_{HZZ}$ at 95\% C.L., where the results are obtained from both current LHC measurements and future colliders analysis, including HL-LHC, CEPC, ILC and CLIC. The other interpretations of these results are given in the appendix~\ref{sec:matching}, including the effective $\cp$-odd fraction $f_{\cp}^{HZZ}$ and the coupling $\Tilde{c}_{ZZ}$.}
    \label{tab:comp_exp}
\end{table}
\section{Conclusions}

This paper mainly discusses the study of $\cp$-properties via the process $e^+ e^-\rightarrow HZ \rightarrow H \mu^- \mu^+$ with a center-of-mass energy 250~GeV and the transversely and longitudinally polarised $e^\pm$ beams at the ILC. In this paper, we carried out an analytical computation of the differential cross section for the Higgs-strahlung process with a $Z$-boson decaying into two muons while incorporating the effects of initial polarisation. Applying full spin correlations, we investigated the impact of $\cp$-violating couplings on the muon azimuthal angular distributions and discovered that the partial cross sections for the regions of $\eta_H \sin2\phi_{\mu^-} > 0$ and $\eta_H \sin2\phi_{\mu^-} < 0$ are asymmetric. Particularly, the azimuthal angle of the muons pair is defined by the orientation of the transverse polarisation of the initial beams. Based on the analysis of angular distributions, we construct a $\cp$-odd observable $\mathcal{O}^T_\cp$ in Eq.~\eqref{eq:obscpo}, which is odd under the naive $\mathcal{T}$ reversal transformation. This $\cp$-odd observable can be used to construct the asymmetry $\mathcal{A}^T_\cp$, which is sensitive to the $\cp$-violation. Based on the analytical calculation, we know that the size of $\mathcal{A}^T_\cp$ is highly depending on the polarisation fraction, and the larger polarisation fraction leads to larger $\mathcal{A}^T_\cp$.

In addition, the other $\cp$-odd observable $\mathcal{O}^{UL}_\cp$ can be constructed as well, which can be always measured whatever initial beams polarisation is applied. The asymmetry $\mathcal{A}^{UL}_\cp$, defined by $\mathcal{O}^{UL}_\cp$, is independent on the polarisation fraction for both longitudinal and transverse polarisation. Since the $\mathcal{O}^{UL}_\cp$ is a different observable as $\mathcal{O}^T_\cp$, one can combine these two observables to increase the statistical significance, when transverse polarisation is imposed. On the other hand, the statistical fluctuation can be suppressed by enhancing total cross-section when longitudinal polarisation is imposed. Therefore, the longitudinal polarisation can be helpful to increase the sensitivity to $\cp$-violation as well.

Furthermore, we performed the Monte-Carlo simulation for these process at 250~GeV center-of-mass energy with initial polarised beams by \texttt{whizard-3.0.3}. For the data generated by MC simulation, we made the fit for the number of events in the signal regions, and obtain the asymmetries by the fitting results. Particularly, we setup two scenarios for varying $\cp$-violation effect. Firstly we vary the $\cp$-mixing angle $\xi_\cp$ with fixing total cross-section. With the help of this scenario, we can determine the limit of intrinsic $\cp$-mixing angle $|\xi_\cp|\sim0.035$ with 5~ab$^{-1}$ and transverse polarisation only $(P^T_{-},P^T_{+})=(90\%, 40\%)$, and $|\xi_\cp|\sim0.03$ with longitudinal polarisation only $(P^L_{-},P^L_{+})=(-90\%, 40\%)$. 
The other scenario is fixing the SM tree-level $HZZ$ interaction and vary the additional contribution from $\cp$-odd coupling $\widetilde{c}_{HZZ}$. In this case, we can determine the limit of $\cp$-odd couplings $\widetilde{c}_{HZZ}\sim 0.011$ with 5~ab$^{-1}$ and transverse or longitudinal polarisation $(|P_{-}|,|P_{+}|)=(90\%, 40\%)$.


By comparing with the other analysis for the $\widetilde{c}_{HZZ}$ coupling, the precision via Higgs strahlung process at $e^+ e^-$ collider 250~GeV can be significantly better than current LHC measurements. Concerning the analysis at CEPC with unpolarised beams, the initial polarised beams can improve the sensitivity to $\cp$-violation effect, for both transverse and longitudinal polarisation. The reason of the improvement is because that the transverse polarisation can provide additional observable, while the longitudinal polarisation can increase the total cross-section and suppress the statistical uncertainty. Additionally, the $Z$-fusion process at 1~TeV CLIC can provide the similar precision of $\cp$-odd coupling to the ILC with polarised beams, but the Higgs strahlung process at ILC has the lower center-of-mass energy. 

Overall, these determination of $\cp$-odd coupling limits is an optimistic estimation, which did not take the full background analysis and systematic uncertainties into account. However, based on these analysis, we have learned which effect contributed by the initial beams polarisation, and obtained a method to improve the sensitivity to $\cp$-violation effect of $HZZ$ interaction, when transverse or longitudinal polarisation is imposed.

\section*{Acknowledgements}
G. Moortgat-Pick acknowledge support by the Deutsche Forschungsgemeinschaft (DFG, German Research Foundation) under Germany’s Excellence Strategy EXC 2121 "Quantum Universe"- 390833306. We thank to J. Reuter and W. Kilian for \texttt{Whizard} technical support. We thank to N. Rehberg and S. Hardt for cross checking and comparison.

\begin{appendix}
 \section{The analytical result of cross section}
\label{sec:xseehmumu}
In order to calculate the cross section of $e^-(p_{e^-}) e^+(p_{e^+}) \rightarrow Z(q_2) H(p_H) \rightarrow \mu^-(p_{\mu^-})\mu^+(p_{\mu^+})$ process, we applied the narrow width approximation, and calculate the Higgs strahlung and $Z$ decay separately.

For the SM Higgs strahlung $e^-(p_{e^-},\lambda_r) e^+(p_{e^+}, \lambda_u) \rightarrow Z(q_2, \lambda^i) H$, the scattering amplitude with the spin indices of the initial electron, positron and the $Z$ boson is given by:
{\fontsize{11}{11}\selectfont
\begin{equation}
    \mathcal{M}_{\lambda_r \lambda_s}^{\lambda^i}=\bar{v}(p_{e^+},\lambda_u)\Big[ i\frac{g^2 m_Z}{2 c_W^2 }\frac{-\eta_{\mu\nu}+\frac{k_\mu k_\nu}{m_Z^2}}{k^2-m_Z^2}\gamma^\mu(c_V+c_A\gamma_5) {\varepsilon^*}^\nu(q_2, \lambda^i)\Big] u(p_{e^-}, \lambda_r).
\end{equation}}

After applying the Bouchiat-Michel formula of Eq.~\eqref{eq:boumi_uu} and polarization matrix of Eq.~\eqref{eq:polmat}, the result of the scattering amplitude square is given by:
\begin{equation}
       \rho^{ii'}= (1-P^3_-P^3_+)A + (P^3_- - P^3_+)B + \sum^{1,2}_{mn}P^m_- P^n_+ C_{mn}.
\end{equation}
For the SM, the unpolarized part is:
\begin{multline}
    A_\text{SM} = \frac{g^4m_Z^2}{4c_W^2(s-m_Z^2)}\Big( (c_V^2 +c_A^2)[(\varepsilon^i\cdot p_{e^+})({\varepsilon^*}^{i'}\cdot p_{e^-})+(\varepsilon^i\cdot p_{e^-})({\varepsilon^*}^{i'}\cdot p_{e^+})-(\varepsilon^i\cdot {\varepsilon^*}^{i'})(p_{e^+}\cdot p_{e^-})]\\-{i}2c_Vc_A\epsilon_{\alpha\mu\beta\nu}{\varepsilon^i}^\alpha{\varepsilon^\beta}^{i'} {p_{e^-}}^\mu {p_{e^+}}^\nu\Big),
\end{multline}
and the longitudinally polarized part is 
\begin{multline}
    B_\text{SM} = \frac{g^4m_Z^2}{4c_W^2(s-m_Z^2)}\Big( 2c_V c_A[(\varepsilon^i\cdot p_{e^+})({\varepsilon^*}^{i'}\cdot p_{e^-})+(\varepsilon^i\cdot p_{e^-})({\varepsilon^*}^{i'}\cdot p_{e^+})-(\varepsilon^i\cdot {\varepsilon^*}^{i'})(p_{e^+}\cdot p_{e^-})]\\-{i}(c_V^2+c_A^2)\epsilon_{\alpha\mu\beta\nu}{\varepsilon^i}^\alpha{\varepsilon^\beta}^{i'} {p_{e^-}}^\mu {p_{e^+}}^\nu\Big),
\end{multline}
as well as the transversely polarized part
{\small
\begin{multline}
    C^{mn}_\text{SM}=\frac{g^4m_Z^2(c_A^2-c_V^2)}{4c_W^2(s-m_Z^2)}\Big(  (s_-^m\cdot s_+^n)[(\varepsilon^i\cdot p_{e^-})({\varepsilon^*}^{i'}\cdot p_{e^+})+(\varepsilon^i\cdot p_{e^+})({\varepsilon^*}^{i'}\cdot p_{e^-})-(\varepsilon^i\cdot{\varepsilon^*}^{i'})(p_{e^-}\cdot p_{e^+})] \\+(p_{e^-}\cdot p_{e^+})[(\varepsilon^i\cdot s_+^n)({\varepsilon^*}^{i'}\cdot s_-^m)+(\varepsilon^i\cdot s_-^m)({\varepsilon^*}^{i'}\cdot s_+^n)] \Big).
\end{multline}}

For the $Z(q_2)\rightarrow \mu^-(p_{\mu^-}) \mu^+(p_{\mu^+})$ decay, we have the amplitude:
\begin{equation}
    \mathcal{M}^i = \frac{ig}{2c_W}{\varepsilon^\lambda}^i  \bar{u}(p_{\mu^-})\gamma_\lambda (c_V+c_A\gamma_5)v(p_{\mu^+}),
\end{equation}
\begin{multline}
    \rho^{i i'}_D = \frac{g^2}{c_W^2}\Big( (c_V^2+c_A^2)[(\varepsilon^i\cdot p_{\mu^-})({\varepsilon^*}^{i'}\cdot p_{\mu^+}) + (\varepsilon^i\cdot p_{\mu^+})({\varepsilon^*}^{i'}\cdot p_{\mu^-})+\delta^{ii'}(p_{\mu^-}\cdot p_{\mu^+})] \\-i2c_V c_A \epsilon_{\alpha\mu\beta\nu}{\varepsilon^i}^\alpha {{\varepsilon^*}^{i'}}^\beta p^\mu_5 p_{\mu^+}^\nu \Big).
\end{multline}

By using the narrow width approximation of Eq.~\eqref{eq:narrowxs}, the full scattering amplitude square is given by:
{
\begin{equation}
    \begin{split}
&|M|^2 = \frac{g^6 m_Z}{4c_W^6(s-m_Z^2)^2\Gamma_Z} \Big\{    (1-P_-^3 P_+^3)\Bigl[ 2(c_V^2+c_A^2)^2\Big(\frac{(p_{e^-}\cdot q_2)(p_{e^+}\cdot q_2)}{m_Z^2}(p_{\mu^-}\cdot p_{\mu^+}) \\&\qquad- \frac{(q_2\cdot p_{\mu^-})^2}{m_Z^2}(p_{e^-}\cdot p_{e^+})-2[\frac{(p_{e^-}\cdot q_2)(q_2\cdot p_{\mu^-})}{m_Z^2}-p_{e^-}\cdot p_{\mu^-}][\frac{(p_{e^+}\cdot q_2)(q_2\cdot p_{\mu^-})}{m_Z^2}-p_{e^+}\cdot p_{\mu^-}]\Big)\\&\qquad-24c_V^2 c_A^2[(p_{e^-}\cdot p_{\mu^-})(p_{e^+}\cdot p_{\mu^+})-(p_{e^-}\cdot p_{\mu^+})(p_{e^+}\cdot p_{\mu^+})] \Bigr]\\
&+(P^3_- -P^3_+)2c_V c_A\Bigl[    (c_V^2+c_A^2)\Big(\frac{(p_{e^-}\cdot q_2)(p_{e^+}\cdot q_2)}{m_Z^2}(p_{\mu^-}\cdot p_{\mu^+}) - \frac{(q_2\cdot p_{\mu^-})^2}{m_Z^2}(p_{e^-}\cdot p_{e^+})\\
&\qquad-2[\frac{(p_{e^-}\cdot q_2)(q_2\cdot p_{\mu^-})}{m_Z^2}-p_{e^-}\cdot p_{\mu^-}][\frac{(p_{e^+}\cdot q_2)(q_2\cdot p_{\mu^-})}{m_Z^2}-p_{e^+}\cdot p_{\mu^-}]\\&\qquad-6[(p_{e^-}\cdot p_{\mu^-})(p_{e^+}\cdot p_{\mu^+})-(p_{e^-}\cdot p_{\mu^+})(p_{e^+}\cdot p_{\mu^+})]\Big)\Bigr]\\
&-\sum_{m,n}P^m_-P^n_+ \Bigl[(c_V^4-c_A^4)\big[(s_-^m\cdot s_+^n)\Bigl(\frac{(p_{e^-}\cdot q_2)(p_{e^+}\cdot q_2)}{m_Z^2}(p_{\mu^-}\cdot p_{\mu^+}) - \frac{(q_2\cdot p_{\mu^-})^2}{m_Z^2}(p_{e^-}\cdot p_{e^+})\\
&\qquad-2[\frac{(p_{e^-}\cdot q_2)(q_2\cdot p_{\mu^-})}{m_Z^2}-p_{e^-}\cdot p_{\mu^-}][\frac{(p_{e^+}\cdot q_2)(q_2\cdot p_{\mu^-})}{m_Z^2}-p_{e^+}\cdot p_{\mu^-}]\Bigr)\\&\qquad+2(p_{e^-}\cdot p_{e^+})\Bigl((p_{\mu^-}\cdot p_{\mu^+})(\frac{(q_2\cdot s_+^n)(q_2\cdot s_-^m)}{m_Z^2}-s_+^n\cdot s_-^m)\\
&\qquad-2(\frac{(q_2\cdot s_+^n)(q_2\cdot p_{\mu^-})}{m_Z^2}-p_{\mu^-}\cdot s_+^n)(\frac{(q_2\cdot s_-^m)(q_2\cdot p_{\mu^-})}{m_Z^2}-p_{\mu^-}\cdot s_-^m)\Bigr)\big]  \Bigr]
\Big\}.
\end{split}
\end{equation}}

For the case with the BSM $\mathcal{CP}$-odd contribution, the result of the full scattering amplitude squared still takes the form of Eq.~\eqref{eq:matpol}, which can be separated into the three parts as well. Therefore, we have:
\begin{multline}
    |\mathcal{M}|^2= (1-P^3_-P^3_+)( c_\alpha^2 \kappa^2_\text{SM} A_\text{SM} + s_\alpha c_\alpha\kappa_\text{SM}\Tilde{\kappa}_{HZZ} {A}_\text{CP-odd} +s_\alpha^2 \Tilde{\kappa}_{HZZ}^2 \tilde{{A}}_\text{CP-even}) \\
    + (P^3_- - P^3_+)( c_\alpha^2 \kappa^2_\text{SM} B_\text{SM} + s_\alpha c_\alpha\kappa_\text{SM}\Tilde{\kappa}_{HZZ} B_{\mathrm{CP-odd}} +s_\alpha^2 \Tilde{\kappa}_{HZZ}^2 \tilde{B}_{\mathrm{CP-even}}) \\
    + \sum^{1,2}_{mn}P^m_- P^n_+ ( c_\alpha^2 \kappa^2_\text{SM} C^{mn}_\text{SM} + s_\alpha c_\alpha\kappa_\text{SM}\Tilde{\kappa}_{HZZ} C^{mn}_{\mathrm{CP-odd}} +s_\alpha^2 \Tilde{\kappa}_{HZZ}^2 C^{mn}_{\mathrm{CP-even}}),
\end{multline}
where for the $\mathcal{CP}$-odd part, we have:
\begin{multline}
     A_{\mathrm{CP-odd}} = - \frac{g^6 \epsilon_{\alpha\beta\mu\nu}p_{e^-}^\alpha p_{e^+}^\beta q_2^\mu p_{\mu^-}^\nu}{c_W^6(s-m_Z^2)^2m_Z^3\Gamma_Z}\Big[ 2(c_V^2 + c_A^2)^2(q_2\cdot p_{\mu^-})(p_{e^-} -p_{e^+})\cdot q_2 \\+m_Z^2 (2c_V^2c_A^2(p_{e^-} + p_{e^+})\cdot q_2-(c_V^2 + c_A^2)^2 (p_{e^-} -p_{e^+} )\cdot p_{\mu^-})\Big],
     \label{eq:cpmixa}
\end{multline}
\begin{multline}
     B_{\mathrm{CP-odd}} = \frac{g^6 \epsilon_{\alpha\beta\mu\nu}p_{e^-}^\alpha p_{e^+}^\beta q_2^\mu p_{\mu^-}^\nu}{c_W^6(s-m_Z^2)^2m_Z^3\Gamma_Z}c_A c_V(c_A^2 + c_V^2)[2(q_2\cdot p_{\mu^-}) (p_{e^-}-p_{e^+})\cdot q_2\\+m_Z^2 ((p_{e^-}+p_{e^+})\cdot q_2 -2 (p_{e^-}-p_{e^+})\cdot p_{\mu^-})]
\end{multline}
and
\begin{multline}
     C^{mn}_{\mathrm{CP-odd}} = -\frac{g^6(c_A^4 - c_V^4)}{2 c_W^6(s-m_Z^2)^2m_Z^3\Gamma_Z}\Big[ 2(s_+^n\cdot s_-^m)((q_2\cdot p_{\mu^-} )(p_{e^-}-p_{e^+})\cdot q_2\\-m_Z^2 (p_{e^-}-p_{e^+})\cdot p_{\mu^-})\epsilon_{\alpha\beta\mu\nu}p_{e^-}^\alpha p_{e^+}^\beta q_2^\mu p_{\mu^-}^\nu \\
     + s[\epsilon_{\alpha\beta\mu\nu}(p_{e^-}+p_{e^+})^\alpha q_2^\beta p_{\mu^-}^\mu {s_-^m}^\nu((q_2\cdot p_{\mu^-})(s_+^n\cdot q_2)-(s_+^n\cdot p_{\mu^-})m_Z^2)  \\+ \epsilon_{\alpha\beta\mu\nu}(p_{e^-}+p_{e^+})^\alpha q_2^\beta p_{\mu^-}^\mu {s_+^n}^\nu((q_2\cdot p_{\mu^-})(s_-^m\cdot q_2)-(s_-^m\cdot p_{\mu^-})m_Z^2)     ]  \Big].
     \label{eq:cpmixc}
\end{multline}

Lastly, the part of the $\widetilde{c}^2_{HZZ}$ contributions are:
{\small
\begin{multline}
\tilde{A}_{\mathrm{CP-even}} = \frac{g^6}{8m_Z^3 c_W^6 (s-m_Z^2)^2\Gamma_Z}\Big[-8(c_A^2 + c_V^2)((p_{e^-}\cdot p_{\mu^-}) (p_{e^+}\cdot q_2) - (p_{e^-}\cdot q_2) (p_{e^+}\cdot p_{\mu^-}))^2\\
    + 2s (q_2\cdot p_{\mu^-}) ((c_A^4+c_V^4)[((p_{e^-}\cdot q_2))^2 + ((p_{e^+}\cdot q_2))^2-2((p_{e^-}+p_{e^+})\cdot q_2)((p_{e^-}-p_{e^+})\cdot p_{\mu^-})]\\
+2c_A^2 c_V^2[3((p_{e^-}\cdot q_2))^2 - ((p_{e^+}\cdot q_2))^2-2((p_{e^-}-p_{e^+})\cdot q_2)((p_{e^-}-p_{e^+})\cdot p_{\mu^-})])
    \\-   s m_Z^2[-2(c_V^2 +c_A^2)^2((p_{e^-}-p_{e^+})\cdot p_{\mu^-})^2  -8c_A^2c_V^2 ((p_{e^-}-p_{e^+})\cdot p_{\mu^-})((p_{e^-}-p_{e^+})\cdot q_2) ]\\ -s^2 m_Z^2 (c_A^2 + c_V^2)^2(q_2\cdot p_{\mu^-})\Big],
\end{multline}}
\begin{multline}
 \tilde{B}_{\mathrm{CP-even}}=\frac{g^6 c_A c_V (c_A^2 + c_V^2)}{8m_Z^3 c_W^6 (s-m_Z^2)^2\Gamma_Z}   \Bigl[8((p_{e^-}\cdot p_{\mu^-}) (p_{e^+}\cdot q_2) - (p_{e^-}\cdot q_2) (p_{e^+}\cdot p_{\mu^-}))^2 \\
  + s^2 m_Z^2 (q_2\cdot p_{\mu^-}) -  4s (q_2\cdot p_{\mu^-})[(p_{e^-}\cdot q_2)^2- ((p_{e^-}-p_{e^+})\cdot q_2)((p_{e^-}-p_{e^+})\cdot p_{\mu^-})]\\
 +2s m_Z^2 [((p_{e^-}-p_{e^+})\cdot q_2)((p_{e^-}+p_{e^+})\cdot p_{\mu^-}) - ((p_{e^-}\cdot p_{\mu^-}) - (p_{e^+}\cdot p_{\mu^-}))]\Bigr]
\end{multline}
and
\begin{multline}
 \tilde{C}^{mn}_{\mathrm{CP-even}}=-\frac{g^6(c_A^4 - c_V^4)}{8m_Z^3 c_W^6 (s-m_Z^2)^2\Gamma_Z}   \Big[8 (s_-^m\cdot s_+^n) [(p_{e^-}\cdot q_2)(p_{e^+}\cdot p_{\mu^-}) -(p_{e^+}\cdot q_2)(p_{e^-}\cdot p_{\mu^-})]^2\\
     + 2s \Bigl((s_-^m\cdot s_+^n) (q_2\cdot p_{\mu^-}) [ (p_{e^-}\cdot q_2) (p_{e^+}\cdot q_2) - (p_{e^-}\cdot p_{\mu^-}) (p_{e^-}\cdot q_2) - (p_{e^+}\cdot q_2) (p_{e^+}\cdot p_{\mu^-})\\ - 3 (p_{e^-}\cdot p_{\mu^-}) (p_{e^+}\cdot q_2)  -3 (p_{e^-}\cdot q_2) (p_{e^+}\cdot p_{\mu^-})  ]\\
+ ((p_{e^-}  + p_{e^+})\cdot q_2) ((p_{e^-}+p_{e^+})\cdot p_{\mu^-}) ((s_-^m\cdot p_{\mu^-}) (s_+^n\cdot q_2)+(s_-^m\cdot q_2) (s_+^n\cdot p_{\mu^-}))\\
- ((p_{e^-}+ p_{e^+})\cdot p_{\mu^-})^2 (s_-^m\cdot q_2) (s_+^n\cdot q_2)- ((p_{e^-}+p_{e^+})\cdot q_2)^2 (s_+^n\cdot p_{\mu^-}) (s_-^m\cdot p_{\mu^-}) 
    \Bigr)\\
+ s m_Z^2 [2(s_-^m\cdot s_+^n)(((p_{e^-} + p_{e^+})\cdot p_{\mu^-} )^2 + 4 (p_{e^-}\cdot p_{\mu^-}) (p_{e^+}\cdot p_{\mu^-}))]\\
    +s^2 (q_2\cdot p_{\mu^-}) [(s_-^m\cdot q_2)(s_+^n\cdot q_2)-2(s_-^m\cdot p_{\mu^-})(s_+^n\cdot q_2)-(s_+^n\cdot p_{\mu^-})(s_-^m\cdot q_2) \\+ 2(q_2\cdot p_{\mu^-})(s_-^m\cdot s_+^n)]
 \Big]
 \Big].
\end{multline}
Note that, the internal $Z$ boson momentum can be converted to the momentum of the Higgs boson by the momentum conservation
\begin{equation}
    q_2 = p_{e^-} + p_{e^+} - p_H.
\end{equation}
The total cross section can be the above result applied into the Eq.~\eqref{eq:xstot}, and obtained by the numerical integration.

\section{The phase space}
\label{sec:dlips}
As we know, one can eventually obtain the cross section of $e^+ e^- \rightarrow H \mu^+ \mu^-$ process by integrating over the three-body phase space. However, one of the degrees of freedom can be integrated out by applying the narrow width approximation, and there are only four degrees of freedom in the final phase space. In this case, the Lorentz invariant phase space is given by:
\begin{equation}
    dQ=\frac{1}{(2\pi)^4}\frac{d\Omega_H d\Omega_{\mu^-}}{16\sqrt{s}}\frac{|p_H^f| |p_{\mu^-}^f|}{|\Vec{p}_H+\Vec{p}_{\mu^-}|+|p_{\mu^-}^f|+|p_H^f| \cos\theta_{H\mu}}=Q~d\Omega_H d\Omega_{\mu^-},
    \label{eq:dlips}
\end{equation}
where:
\begin{align}
    &|p_H^f| = \frac{1}{2\sqrt{s}}\sqrt{(s-(m_H+m_Z)^2)(s-(m_H-m_Z)^2)},\\
    &|p_{\mu^-}^f| = \frac{m_Z^2}{2\sqrt{|p_H^f|^2 + m_Z^2} + |p_H^f|\cos\theta_{H\mu}}.
\end{align}
The term $\cos\theta_{H\mu}$ indicates the projection of the muon momentum on the Higgs momentum.

In particular, we can evaluate the phase space in the center of mass frame.
If the electron and positron beams are transversely polarized, their spin vector $\Vec{s}_{e^\pm}$ would be perpendicular to their momentum $\Vec{p}_{e^\pm}$. In this case, we can define a coordinate system by using the spin vector and momentum of electron beams $\Vec{s}_{e^-},~\Vec{p}_{e^-}$, where the momentum of final state particles are shown in Fig.~\ref{fig:eehmumu_coord}. Consequently, the projection $\cos\theta_{H\mu}$ can be expressed as :
\begin{equation}
    \begin{split}
        \cos\theta_{H\mu} = &-\sin\theta_H\cos\phi_H \sin\theta_{\mu^-} \cos\phi_{\mu^-}-\sin\theta_H\sin\phi_H \sin\theta_{\mu^-}\sin\phi_{\mu^-}\\
        &-\cos\theta_H \cos\theta_{\mu^-}.
    \end{split}
\end{equation}

\section{Matching relations between different interpretations}
\label{sec:matching}
\subsection*{Effective $\cp$-odd fraction}
In order to test the $\cp$ properties of the Higgs boson, one can define an effective $\cp$-odd fraction $f^{HZZ}_{CP}$,  referring to \cite{Dawson:2013bba}
\begin{equation}
    f^{HZZ}_{\CP} = \frac{\Gamma^{\cp-\mathrm{odd}}_{H\rightarrow ZZ}}{\Gamma^{\cp-\mathrm{even}}_{H\rightarrow ZZ} + \Gamma^{\cp-\mathrm{odd}}_{H\rightarrow ZZ}},
    \label{eq:fcp}
\end{equation}
where $\Gamma^{\cp-\mathrm{odd}}_{H\rightarrow ZZ}$ is the decay width obtained by setting $c_\mathrm{SM}=c_{HZZ}=0$ and $\widetilde{c}_{HZZ}=1$. If we assume that the $\cp$-odd term $\widetilde{c}_{HZZ}$ is the unique BSM contribution, and the SM tree-level contribution stays invariant, the effective $\cp$-odd fraction is given by
\begin{equation}
    f^{HZZ}_{\CP} = 1/\left(1 + \frac{1}{|\widetilde{c}_{HZZ}|^2  \frac{\Gamma^{\cp-\mathrm{odd}}_{H\rightarrow ZZ}}{\Gamma^{\cp-\mathrm{even}}_{H\rightarrow ZZ}}}\right). 
    \label{eq:fcp2}
\end{equation}
where the decay width ratio can be approximately the same as the cross-section ratio, since the branching ratio of $H\rightarrow Z Z $ is small and the contribution of total width by $\cp$-odd $HZZ$ coupling can be negligible. There we can obtain the decay width ratio by
\begin{equation}
    \frac{\Gamma^{\cp-\mathrm{odd}}_{H\rightarrow ZZ}}{\Gamma^{\cp-\mathrm{even}}_{H\rightarrow ZZ}}\sim \frac{\sigma_3}{\sigma_\mathrm{SM}}[pp\rightarrow H\rightarrow 4\ell (13~\mathrm{TeV})] \sim 0.153.
\end{equation}
Since the $f^{HZZ}_{\CP}$ is defined in the Higgs boson decay, the $f^{HZZ}_{\CP}$ is also an unique process independent quantity. Consequently, we can match all the results in Tab.~\ref{tab:comp_exp} to the $f^{HZZ}_{\CP}$ interpretation, which is presented in Tab.~\ref{tab:comp_exp_fcb}
\begin{table}[h]
    \centering
    \resizebox{\linewidth}{!}{\begin{tabular}{c|ccccccc}
    \hline
       Experiments&  ATLAS\cite{ATLAS:2023mqy}& CMS\cite{CMS:2021nnc}& HL-LHC\cite{Cepeda:2019klc}& CEPC\cite{Sha:2022bkt}& CLIC\cite{Karadeniz:2019upm}& CLIC \cite{Vukasinovic:2023jxd,BozovicJelisavcic:2024czi}& ILC\\
       Processes& $H\rightarrow 4\ell$& $H\rightarrow 4\ell$& $H\rightarrow 4\ell$& $ H Z$&  $W$-fusion& $Z$-fusion &$ H Z,~Z\rightarrow \mu^+\mu^-$\\
       $\sqrt{s}$ [GeV]&  13000&  13000& 14000& 240& 3000& 1000& 250\\
       Luminosity~[fb$^{-1}$]&  139& 137& 3000& 5600& 5000& 8000& 5000\\
       $(|P_{-}|,|P_{+}|)$& & & & & & & $(90\%, 40\%)$\\
       \hline
       $f^{HZZ}_{\cp} (\times 10^{-5})$ &&&&&&\\
       95\% C.L. (2$\sigma$)limit &   [-409.82, 873.58]& [-123.78, 74.91]& [-126.54, 126.54]& [-3.92, 3.92]& [-16.66, 16.66]& [-1.85, 1.85]&  [-1.85, 1.53] \\
       \hline
    \end{tabular}}
    \caption{Summary of the limits of $f^{HZZ}_{\CP}$ at 95\% C.L., where the results are obtained from both current LHC measurements and future colliders analysis, including HL-LHC, CEPC, ILC and CLIC.}
    \label{tab:comp_exp_fcb}
\end{table}

Furthermore, one can also define a effective $\cp$-mixing angle $\psi_\cp$ by using the effective $\cp$-odd fraction, which can be extracted by:
\begin{equation}
    \sin^2\psi_\cp = f^{HZZ}_{CP}.
\end{equation}

\subsection*{$\cp$-odd couplings}

The coupling $a_3$ in \cite{CMS:2019ekd} can be converted to another interpretation of $\cp$-odd coupling by the following relation \cite{CMS:2021nnc}
\begin{equation}
    \Tilde{c}_{ZZ} = -\frac{\sin^2\theta_W\cos^2\theta_W}{2\pi\alpha} a_3,
\end{equation}
where the coupling $\Tilde{c}_{ZZ}$ is defined by the following effective Lagrangian in Eq.(21) of \cite{Cepeda:2019klc}
\begin{equation}
    \mathcal{L}_\mathrm{eff} = \frac{g_1^2+g_2^2}{4}\Tilde{c}_{ZZ} \frac{H}{v}Z_{\mu\nu}\Tilde{Z}^{\mu\nu}.
\end{equation}
Therefore, we have the matching relation
\begin{equation}
    \widetilde{c}_{HZZ}=\frac{g_1^2+g_2^2}{4}\Tilde{c}_{ZZ} = \frac{m_Z^2}{v^2} \Tilde{c}_{ZZ}.
\end{equation}

By using the matching relation, we can convert our results to the $\tilde{c}_{ZZ}$ interpretation, and the summary table of $\tilde{c}_{ZZ}$ is given by Tab.~\ref{tab:comp_exp_czz}.
\begin{table}[h]
    \centering
    \resizebox{\linewidth}{!}{\begin{tabular}{c|ccccccc}
    \hline
       Experiments&  ATLAS\cite{ATLAS:2023mqy}& CMS\cite{CMS:2021nnc}& HL-LHC\cite{Cepeda:2019klc}& CEPC\cite{Sha:2022bkt}& CLIC\cite{Karadeniz:2019upm}& CLIC \cite{Vukasinovic:2023jxd,BozovicJelisavcic:2024czi}& ILC\\
       Processes& $H\rightarrow 4\ell$& $H\rightarrow 4\ell$& $H\rightarrow 4\ell$& $ H Z$&  $W$-fusion& $Z$-fusion &$ H Z,~Z\rightarrow \mu^+\mu^-$\\
       $\sqrt{s}$ [GeV]&  13000&  13000& 14000& 240& 3000& 1000& 250\\
       Luminosity~[fb$^{-1}$]&  139& 137& 3000& 5600& 5000& 8000& 5000\\
       $(|P_{-}|,|P_{+}|)$& & & & & & & $(90\%, 40\%)$\\
       \hline
       $\tilde{c}_{ZZ}$ &&&&&&\\
       95\% C.L. (2$\sigma$)limit&   [-1.2, 1.75]& [-0.66, 0.51]& [-0.66, 0.66]& [-0.12, 0.12]& [-0.24, 0.24]& [-0.08, 0.08]&  [-0.08, 0.07] \\
       \hline
    \end{tabular}}
    \caption{Summary of the limits of $\tilde{c}_{ZZ}$ at 95\% C.L., where the results are obtained from both current LHC measurements and future colliders analysis, including HL-LHC, CEPC, ILC and CLIC.}
    \label{tab:comp_exp_czz}
\end{table}   
\end{appendix}

\bibliographystyle{JHEP}   
\bibliography{reference}

\providecommand{\href}[2]{#2}\begingroup\raggedright\begin{thebibliography}{10}

\bibitem{ATLAS:2012yve}
{\bf ATLAS} Collaboration, G.~Aad et~al., {\it {Observation of a new particle
  in the search for the Standard Model Higgs boson with the ATLAS detector at
  the LHC}},  {\em Phys. Lett. B} {\bf 716} (2012) 1--29,
  [\href{http://arxiv.org//abs/1207.7214}{{\tt arXiv:1207.7214}}].

\bibitem{CMS:2012qbp}
{\bf CMS} Collaboration, S.~Chatrchyan et~al., {\it {Observation of a New Boson
  at a Mass of 125 GeV with the CMS Experiment at the LHC}},  {\em Phys. Lett.
  B} {\bf 716} (2012) 30--61, [\href{http://arxiv.org//abs/1207.7235}{{\tt
  arXiv:1207.7235}}].

\bibitem{Planck:2018vyg}
{\bf Planck} Collaboration, N.~Aghanim et~al., {\it {Planck 2018 results. VI.
  Cosmological parameters}},  {\em Astron. Astrophys.} {\bf 641} (2020) A6,
  [\href{http://arxiv.org//abs/1807.06209}{{\tt arXiv:1807.06209}}]. [Erratum:
  Astron.Astrophys. 652, C4 (2021)].

\bibitem{Sakharov:1967dj}
A.~D. Sakharov, {\it {Violation of CP Invariance, C asymmetry, and baryon
  asymmetry of the universe}},  {\em Pisma Zh. Eksp. Teor. Fiz.} {\bf 5} (1967)
  32--35.

\bibitem{Lee:1973iz}
T.~D. Lee, {\it {A Theory of Spontaneous T Violation}},  {\em Phys. Rev. D}
  {\bf 8} (1973) 1226--1239.

\bibitem{Gunion:2005ja}
J.~F. Gunion and H.~E. Haber, {\it {Conditions for CP-violation in the general
  two-Higgs-doublet model}},  {\em Phys. Rev. D} {\bf 72} (2005) 095002,
  [\href{http://arxiv.org//abs/hep-ph/0506227}{{\tt hep-ph/0506227}}].

\bibitem{ATLAS:2019nvo}
{\bf ATLAS} Collaboration, {\it {Analysis of $t\bar{t}H$ and $t\bar{t}W$
  production in multilepton final states with the ATLAS detector}}, .

\bibitem{ATLAS:2020pvn}
{\bf ATLAS} Collaboration, {\it {Measurement of the properties of Higgs boson
  production at $\sqrt{s}$=13 TeV in the $H\rightarrow \gamma\gamma$ channel
  using 139~{fb}$^{-1}$ of $pp$ collision data with the ATLAS experiment}}, .

\bibitem{ATLAS:2020ior}
{\bf ATLAS} Collaboration, G.~Aad et~al., {\it {$CP$ Properties of Higgs Boson
  Interactions with Top Quarks in the $t\bar{t}H$ and $tH$ Processes Using $H
  \rightarrow \gamma\gamma$ with the ATLAS Detector}},  {\em Phys. Rev. Lett.}
  {\bf 125} (2020), no.~6 061802, [\href{http://arxiv.org//abs/2004.04545}{{\tt
  arXiv:2004.04545}}].

\bibitem{CMS:2019lcn}
{\bf CMS} Collaboration, {\it {Measurement of $\mathrm{t\overline{t}H}$
  production in the $\mathrm{H\rightarrow b\overline{b}}$ decay channel in
  $41.5\,\mathrm{fb}^{-1}$ of proton-proton collision data at
  $\sqrt{s}=13\,\mathrm{TeV}$}}, .

\bibitem{CMS:2020cga}
{\bf CMS} Collaboration, A.~M. Sirunyan et~al., {\it {Measurements of
  $\mathrm{t\bar{t}}H$ Production and the CP Structure of the Yukawa
  Interaction between the Higgs Boson and Top Quark in the Diphoton Decay
  Channel}},  {\em Phys. Rev. Lett.} {\bf 125} (2020), no.~6 061801,
  [\href{http://arxiv.org//abs/2003.10866}{{\tt arXiv:2003.10866}}].

\bibitem{Bahl:2020wee}
H.~Bahl, P.~Bechtle, S.~Heinemeyer, J.~Katzy, T.~Klingl, K.~Peters,
  M.~Saimpert, T.~Stefaniak, and G.~Weiglein, {\it {Indirect $\mathcal{CP}$
  probes of the Higgs-top-quark interaction: current LHC constraints and future
  opportunities}},  {\em JHEP} {\bf 11} (2020) 127,
  [\href{http://arxiv.org//abs/2007.08542}{{\tt arXiv:2007.08542}}].

\bibitem{Gritsan:2022php}
A.~V. Gritsan et~al., {\it {Snowmass White Paper: Prospects of CP-violation
  measurements with the Higgs boson at future experiments}},
  \href{http://arxiv.org//abs/2205.07715}{{\tt arXiv:2205.07715}}.

\bibitem{Dawson:2022zbb}
S.~Dawson et~al., {\it {Report of the Topical Group on Higgs Physics for
  Snowmass 2021: The Case for Precision Higgs Physics}},  in {\em {Snowmass
  2021}}, 9, 2022.
\newblock \href{http://arxiv.org//abs/2209.07510}{{\tt arXiv:2209.07510}}.

\bibitem{Bahl:2022yrs}
H.~Bahl, E.~Fuchs, S.~Heinemeyer, J.~Katzy, M.~Menen, K.~Peters, M.~Saimpert,
  and G.~Weiglein, {\it {Constraining the ${\mathcal {C}}{\mathcal {P}}$
  structure of Higgs-fermion couplings with a global LHC fit, the electron EDM
  and baryogenesis}},  {\em Eur. Phys. J. C} {\bf 82} (2022), no.~7 604,
  [\href{http://arxiv.org//abs/2202.11753}{{\tt arXiv:2202.11753}}].

\bibitem{CMS:2017len}
{\bf CMS} Collaboration, A.~M. Sirunyan et~al., {\it {Constraints on anomalous
  Higgs boson couplings using production and decay information in the
  four-lepton final state}},  {\em Phys. Lett. B} {\bf 775} (2017) 1--24,
  [\href{http://arxiv.org//abs/1707.00541}{{\tt arXiv:1707.00541}}].

\bibitem{CMS:2019jdw}
{\bf CMS} Collaboration, A.~M. Sirunyan et~al., {\it {Constraints on anomalous
  $HVV$ couplings from the production of Higgs bosons decaying to $\tau$ lepton
  pairs}},  {\em Phys. Rev. D} {\bf 100} (2019), no.~11 112002,
  [\href{http://arxiv.org//abs/1903.06973}{{\tt arXiv:1903.06973}}].

\bibitem{CMS:2019ekd}
{\bf CMS} Collaboration, A.~M. Sirunyan et~al., {\it {Measurements of the Higgs
  boson width and anomalous $HVV$ couplings from on-shell and off-shell
  production in the four-lepton final state}},  {\em Phys. Rev. D} {\bf 99}
  (2019), no.~11 112003, [\href{http://arxiv.org//abs/1901.00174}{{\tt
  arXiv:1901.00174}}].

\bibitem{CMS:2021nnc}
{\bf CMS} Collaboration, A.~M. Sirunyan et~al., {\it {Constraints on anomalous
  Higgs boson couplings to vector bosons and fermions in its production and
  decay using the four-lepton final state}},  {\em Phys. Rev. D} {\bf 104}
  (2021), no.~5 052004, [\href{http://arxiv.org//abs/2104.12152}{{\tt
  arXiv:2104.12152}}].

\bibitem{CMS:2022uox}
{\bf CMS} Collaboration, A.~Tumasyan et~al., {\it {Constraints on anomalous
  Higgs boson couplings to vector bosons and fermions from the production of
  Higgs bosons using the \ensuremath{\tau}\ensuremath{\tau} final state}},
  {\em Phys. Rev. D} {\bf 108} (2023), no.~3 032013,
  [\href{http://arxiv.org//abs/2205.05120}{{\tt arXiv:2205.05120}}].

\bibitem{CMS:2024bua}
{\bf CMS} Collaboration, A.~Hayrapetyan et~al., {\it {Constraints on anomalous
  Higgs boson couplings from its production and decay using the WW channel in
  proton-proton collisions at $\sqrt{s}$ = 13 TeV}},
  \href{http://arxiv.org//abs/2403.00657}{{\tt arXiv:2403.00657}}.

\bibitem{ATLAS:2020evk}
{\bf ATLAS} Collaboration, G.~Aad et~al., {\it {Test of CP invariance in
  vector-boson fusion production of the Higgs boson in the
  $H\rightarrow\tau{\tau}$ channel in proton-proton collisions at s=13TeV with
  the ATLAS detector}},  {\em Phys. Lett. B} {\bf 805} (2020) 135426,
  [\href{http://arxiv.org//abs/2002.05315}{{\tt arXiv:2002.05315}}].

\bibitem{ATLAS:2022tan}
{\bf ATLAS} Collaboration, G.~Aad et~al., {\it {Test of CP Invariance in Higgs
  Boson Vector-Boson-Fusion Production Using the $H\rightarrow{\gamma}{\gamma}$
  Channel with the ATLAS Detector}},  {\em Phys. Rev. Lett.} {\bf 131} (2023),
  no.~6 061802, [\href{http://arxiv.org//abs/2208.02338}{{\tt
  arXiv:2208.02338}}].

\bibitem{ATLAS:2023mqy}
{\bf ATLAS} Collaboration, G.~Aad et~al., {\it {Test of CP-invariance of the
  Higgs boson in vector-boson fusion production and its decay into four
  leptons}},  \href{http://arxiv.org//abs/2304.09612}{{\tt arXiv:2304.09612}}.

\bibitem{Cepeda:2019klc}
M.~Cepeda et~al., {\it {Report from Working Group 2}: {Higgs Physics at the
  HL-LHC and HE-LHC}},  {\em CERN Yellow Rep. Monogr.} {\bf 7} (2019) 221--584,
  [\href{http://arxiv.org//abs/1902.00134}{{\tt arXiv:1902.00134}}].

\bibitem{Ge:2020mcl}
S.-F. Ge, G.~Li, P.~Pasquini, and M.~J. Ramsey-Musolf, {\it {CP-violating Higgs
  Di-tau Decays: Baryogenesis and Higgs Factories}},  {\em Phys. Rev. D} {\bf
  103} (2021), no.~9 095027, [\href{http://arxiv.org//abs/2012.13922}{{\tt
  arXiv:2012.13922}}].

\bibitem{Jovin:2021qnz}
T.~A. Jovin, I.~B. Jelisavcic, I.~Smiljanic, G.~Kacarevic, N.~Vukasinovic,
  G.~M. Dumbelovic, J.~Stevanovic, M.~Radulovic, and D.~Jeans, {\it {Probing
  the CP properties of the Higgs sector at ILC}},  in {\em {International
  Workshop on Future Linear Colliders}}, 5, 2021.
\newblock \href{http://arxiv.org//abs/2105.06530}{{\tt arXiv:2105.06530}}.

\bibitem{Jeans:2018anq}
D.~Jeans and G.~W. Wilson, {\it {Measuring the CP state of tau lepton pairs
  from Higgs decay at the ILC}},  {\em Phys. Rev. D} {\bf 98} (2018), no.~1
  013007, [\href{http://arxiv.org//abs/1804.01241}{{\tt arXiv:1804.01241}}].

\bibitem{Sha:2022bkt}
Q.~Sha et~al., {\it {Probing Higgs CP properties at the CEPC in the $e^{+}
  e^{-} \rightarrow Z H \rightarrow l^{+} l^{-}H$ using optimal variables}},
  {\em Eur. Phys. J. C} {\bf 82} (2022), no.~11 981,
  [\href{http://arxiv.org//abs/2203.11707}{{\tt arXiv:2203.11707}}]. [Erratum:
  Eur.Phys.J.C 83, 62 (2023)].

\bibitem{Karadeniz:2019upm}
O.~Karadeniz, A.~Senol, K.~Y. Oyulmaz, and H.~Denizli, {\it {CP-violating
  Higgs-gauge boson couplings in $H\nu \bar{\nu}$ production at three energy
  stages of CLIC}},  {\em Eur. Phys. J. C} {\bf 80} (2020), no.~3 229,
  [\href{http://arxiv.org//abs/1909.08032}{{\tt arXiv:1909.08032}}].

\bibitem{Vukasinovic:2023jxd}
N.~Vuka\v{s}inovi\'c, I.~Bo\v{z}ovi\'c-Jelisav\v{c}i\'c, and
  G.~Ka\v{c}arevi\'c, {\it {Measurement of the CPV Higgs mixing angle in
  ZZ-fusion at 1 TeV ILC}},  in {\em {International Workshop on Future Linear
  Colliders}}, 7, 2023.
\newblock \href{http://arxiv.org//abs/2307.16514}{{\tt arXiv:2307.16514}}.

\bibitem{Moortgat-Pick:2005jsx}
G.~Moortgat-Pick et~al., {\it {The Role of polarized positrons and electrons in
  revealing fundamental interactions at the linear collider}},  {\em Phys.
  Rept.} {\bf 460} (2008) 131--243,
  [\href{http://arxiv.org//abs/hep-ph/0507011}{{\tt hep-ph/0507011}}].

\bibitem{Rao:2007ce}
K.~Rao and S.~D. Rindani, {\it {Charged lepton distributions as a probe of
  contact e+e-HZ interactions at a linear collider with polarized beams}},
  {\em Phys. Rev. D} {\bf 77} (2008) 015009,
  [\href{http://arxiv.org//abs/0709.2591}{{\tt arXiv:0709.2591}}]. [Erratum:
  Phys.Rev.D 80, 019901 (2009)].

\bibitem{Biswal:2009ar}
S.~S. Biswal and R.~M. Godbole, {\it {Use of transverse beam polarization to
  probe anomalous VVH interactions at a Linear Collider}},  {\em Phys. Lett. B}
  {\bf 680} (2009) 81--87, [\href{http://arxiv.org//abs/0906.5471}{{\tt
  arXiv:0906.5471}}].

\bibitem{Rindani:2010pi}
S.~D. Rindani and P.~Sharma, {\it {Decay-lepton correlations as probes of
  anomalous ZZH and gammaZH interactions in e+e- --\ensuremath{>} ZH with
  polarized beams}},  {\em Phys. Lett. B} {\bf 693} (2010) 134--139,
  [\href{http://arxiv.org//abs/1001.4931}{{\tt arXiv:1001.4931}}].

\bibitem{Kilian:2007gr}
W.~Kilian, T.~Ohl, and J.~Reuter, {\it {WHIZARD: Simulating Multi-Particle
  Processes at LHC and ILC}},  {\em Eur. Phys. J. C} {\bf 71} (2011) 1742,
  [\href{http://arxiv.org//abs/0708.4233}{{\tt arXiv:0708.4233}}].

\bibitem{Moretti:2001zz}
M.~Moretti, T.~Ohl, and J.~Reuter, {\it {O'Mega: An Optimizing matrix element
  generator}},  \href{http://arxiv.org//abs/hep-ph/0102195}{{\tt
  hep-ph/0102195}}.

\bibitem{Artoisenet:2013puc}
P.~Artoisenet et~al., {\it {A framework for Higgs characterisation}},  {\em
  JHEP} {\bf 11} (2013) 043, [\href{http://arxiv.org//abs/1306.6464}{{\tt
  arXiv:1306.6464}}].

\bibitem{Bouchiat:1958yui}
C.~Bouchiat and L.~Michel, {\it {Mesure de la polarisation des electrons
  relativistes}},  {\em Nucl. Phys.} {\bf 5} (1958) 416--434.

\bibitem{Diehl:2003qz}
M.~Diehl, O.~Nachtmann, and F.~Nagel, {\it {Probing triple gauge couplings with
  transverse beam polarisation in e+ e- ---\ensuremath{>} W+ W-}},  {\em Eur.
  Phys. J. C} {\bf 32} (2003) 17--27,
  [\href{http://arxiv.org//abs/hep-ph/0306247}{{\tt hep-ph/0306247}}].

\bibitem{Fleischer:1993ix}
J.~Fleischer, K.~Kolodziej, and F.~Jegerlehner, {\it {Transverse versus
  longitudinal polarization effects in e+ e- ---\ensuremath{>} W+ W-}},  {\em
  Phys. Rev. D} {\bf 49} (1994) 2174--2187.

\bibitem{BozovicJelisavcic:2024czi}
{\bf ILD concept group} Collaboration, I.~Bozovi\'c~Jelisav\v{c}i\'c,
  N.~Vukasinovic, and G.~Kacarevic, {\it {Probing CPV mixing in the Higgs
  sector in VBF at 1 TeV ILC}},  {\em PoS} {\bf EPS-HEP2023} (2024) 404.

\bibitem{Dawson:2013bba}
S.~Dawson et~al., {\it {Working Group Report: Higgs Boson}},  in {\em {Snowmass
  2013}: {Snowmass on the Mississippi}}, 10, 2013.
\newblock \href{http://arxiv.org//abs/1310.8361}{{\tt arXiv:1310.8361}}.

\end{thebibliography}\endgroup
\end{document}